\documentclass[aps,prd,twocolumn,superscriptaddress,preprintnumbers,floatfix,longbibliography,nofootinbib]{revtex4-1}

\usepackage{graphicx}
\usepackage{amsmath}
\usepackage[caption=false]{subfig}
\usepackage{siunitx}
\usepackage{placeins}
\usepackage{color}
\usepackage{standalone}
\usepackage{dcolumn}
\usepackage{bm}
\usepackage{microtype}
\usepackage{etoolbox}
\usepackage{amssymb}
\usepackage{mathrsfs}
\usepackage{accents}
\usepackage{gensymb}
\usepackage{float}
\usepackage[normalem]{ulem}
\usepackage[dvipsnames]{xcolor}
\usepackage[colorlinks,urlcolor=NavyBlue,citecolor=NavyBlue,linkcolor=NavyBlue,pdfusetitle]{hyperref}
\usepackage{subfig}

\usepackage{etoolbox}
\usepackage{xcolor}

\newtoggle{commentsoff}
\iftoggle{commentsoff}{
	\newcommand{\dhj}[1]{}
	\newcommand{\ls}[1]{}
}{
\newcommand{\dhj}[1]{{\color{blue}~\textsf{[{\bf Dana}: #1]}}}
\newcommand{\ls}[1]{{\color{red}~\textsf{[{\bf Lilli}: #1]}}}

}
\newcommand{\hpeak}{h_0^{\rm peak}}
\newcommand{\fpeak}{f_{\rm GW}^{\rm peak}}

\begin{document}

\title{Methods and prospects for gravitational wave searches targeting ultralight \\vector boson clouds around known black holes}

\author{Dana Jones}
\affiliation{OzGrav-ANU, Centre for Gravitational Astrophysics, College of Science, The Australian National University, Australian Capital Territory 2601, Australia}

\author{Ling Sun}
\affiliation{OzGrav-ANU, Centre for Gravitational Astrophysics, College of Science, The Australian National University, Australian Capital Territory 2601, Australia}

\author{Nils Siemonsen}
\affiliation{Perimeter Institute for Theoretical Physics, Waterloo, Ontario N2L 2Y5, Canada}
\affiliation{Arthur B. McDonald Canadian Astroparticle Physics Research Institute, 64 Bader Lane, Queen's University, Kingston, ON K7L 3N6, Canada}
\affiliation{Department of Physics \& Astronomy, University of Waterloo, Waterloo, ON N2L 3G1, Canada}

\author{William E. East}
\affiliation{Perimeter Institute for Theoretical Physics, Waterloo, Ontario N2L 2Y5, Canada}

\author{Susan M. Scott}
\affiliation{OzGrav-ANU, Centre for Gravitational Astrophysics, College of Science, The Australian National University, Australian Capital Territory 2601, Australia}

\author{Karl Wette}
\affiliation{OzGrav-ANU, Centre for Gravitational Astrophysics, College of Science, The Australian National University, Australian Capital Territory 2601, Australia}

\date{\today}

\begin{abstract}
Ultralight bosons are predicted in many extensions to the Standard Model 
and are popular dark matter candidates. The black hole superradiance mechanism 
allows for these particles to be probed
using only their gravitational 
interaction. In this scenario, an ultralight boson cloud
may form spontaneously around a spinning black hole and extract a non-negligible 
fraction of the black hole's mass. These oscillating clouds
produce quasi-monochromatic, long-duration gravitational waves that may be
detectable by ground-based or space-based gravitational wave detectors.
We discuss the capability
of a new long-duration signal tracking method, based on a hidden Markov model, to
detect gravitational wave signals generated by ultralight vector
boson clouds, including cases where the signal frequency evolution timescale is much shorter than that of a typical continuous wave signal.
We quantify the detection horizon distances for vector boson clouds with current- and next-generation ground-based detectors. 
We demonstrate that vector clouds hosted by black holes with mass $\gtrsim 60 M_{\odot}$ and spin $\gtrsim 0.6$ are within the reach of current-generation detectors up to a luminosity distance of $\sim 1$~Gpc. 
This search method enables one to target vector boson clouds around remnant black holes from compact binary mergers detected by gravitational-wave detectors. We discuss the impact of the sky localization of the merger events and demonstrate that a typical remnant black hole reasonably well-localized by the current generation detector network is accessible in a follow-up search.
\end{abstract}

\maketitle

\section{Introduction}
\label{sec:intro}

The first direct detection of gravitational waves (GWs) made by the Advanced Laser Interferometer Gravitational-Wave Observatory (aLIGO) in 2015 ushered in a new and exciting era of astrophysics~\cite{first_detection, aLIGO}. The addition of two detectors, Advanced Virgo~\cite{aVirgo} and KAGRA~\cite{KAGRA}, coupled with continuous upgrades in sensitivity, has led to a total of 90 direct observations of compact binary coalescence (CBC) events to date, with this number expected to grow exponentially in the coming years~\cite{CBCs_O1O2, CBCs_O3a, CBCs_O3b}. In the wake of these discoveries, one of the most exciting prospects is to use GWs to address questions in fundamental physics. From a particle physics perspective, GW detectors are invaluable and unique tools in the search for physics beyond the Standard Model. Specifically, the black hole superradiance mechanism~\cite{Zel'Dovich1971, Misner1972, Starobinskii1973, Detweiler1980, Brito2020} and the resulting detectable gravitational radiation are ideal probes of weakly coupled ultralight bosons~\cite{Arvanitaki2011, Arvanitaki2015} in regions of the parameter space inaccessible to current terrestrial experiments.

Ultralight bosons have been invoked in a variety of settings in order to address open problems in particle physics and cosmology. These include scalar (spin-0), vector (spin-1) particles, as well as massive tensor (spin-2) fields. 
The QCD axion, as well as axion-like particles, are well-motivated ultralight scalar particles that solve the strong CP-problem and may constitute a significant fraction of dark matter~\cite{Peccei_Quinn1977, Peccei_Quinn1977_2, Weinberg1978, Arvanitaki2010}. Similarly, ultralight vector bosons emerge in low-energy limits of quantum gravity models and could contribute to the dark matter density~\cite{Goodsell2009, Holdom1986, Jaeckel2010, Essig2013, Hui2017, Agrawal2020, Fabbrichesi2021}, and general relativity may be modified by massive spin-2 fields~\cite{Clifton:2011jh, Dias:2023ynv}. Typical strategies used in lab experiments to search for these elusive particles rely on weak but non-zero couplings to the Standard Model. The superradiance mechanism around spinning black holes, on the other hand, results in observable smoking gun signatures of the presence of ultralight bosons that depend \textit{only} on gravitational interactions.

Ultralight bosons can form bound states around spinning black holes, growing into macroscopic clouds that produce distinct observational signatures as a result of a time-varying quadrupole moment (and higher moments)~\cite{Arvanitaki2011, Yoshino2014, Yoshino2015, Arvanitaki2015, Arvanitaki2017, Brito2017, Brito2017_2, Yoshino2014, Yoshino2015, Baryakhtar2017, Chan2022, Cardoso2018, Baumann2019, Hannuksela2019, Zhang2019, East2017, East2017_2} (see Ref.~\cite{Brito2020} for a review). As the cloud extracts energy and angular momentum from the black hole through the superradiance mechanism, its amplitude grows exponentially~\cite{Penrose1969, Press_Teukolsky1972, Zel'Dovich1971, Starobinskii1973, Detweiler1980, Bekenstein1973, Dolan2007, Arvanitaki2011}. Considering only gravitational interactions of the ultralight bosons, this unstable behavior saturates due to the spin down of the black hole, resulting in a superradiant cloud that dissipates through GW emission~\cite{East2017_2, East2018}. The subsequent GW emission is quasi-monochromatic and occurs at roughly twice the oscillation frequency of the boson cloud. In particular, the frequency of gravitational radiation from clouds around stellar mass black holes falls squarely within the sensitive band of ground-based GW detectors if the boson mass lies within a range of \mbox{$\sim 10^{-14}$--$10^{-11}$~eV}~\cite{Arvanitaki2015, Brito2017_2}. These GW signals can be searched for in data collected by current and future detector networks, and confident constraints may be placed on the existence of ultralight bosons in the absence of a signal. Black hole spin measurements have previously been used to place constraints on the existence of ultralight scalar~\cite{Arvanitaki2015, Ng2021, Cardoso2018, Brito2017_2} and vector bosons~\cite{Baryakhtar2017, Cardoso2018}, but with significant associated uncertainties.

Various ultralight boson search strategies and target signals have been considered, broadly classified into continuous wave (CW) searches (blind or directed), stochastic GW background searches, and follow-up searches for quasi-continuous GWs from previously observed binary black hole merger events. All-sky searches for scalar bosons using CW search techniques are described in Refs.~\cite{O3_all-sky_scalar_bosons, Palomba2019, Dergachev2019}, studies targeting galactic sources are carried out in Refs.~\cite{KAGRA:2022osp, Zhu:2020tht}, and a directed search of the x-ray binary system Cygnus X-1 is presented in Ref.~\cite{O2_CygnusX1_scalar_bosons}. Searches for a stochastic GW background from a population of black holes with scalar~\cite{Tsukada2019} and vector~\cite{Tsukada2021} boson clouds have been used to place constraints on the respective mass ranges. However, these constraints are subject to assumptions about the underlying black hole population and the past astrophysical history of specific black holes. Follow-up searches targeting remnant black holes formed in binary mergers remedy this shortcoming, as the entire history of the newly formed black hole, as well as its properties, are well understood.

Follow-up searches of merger remnants are therefore ideal for two reasons: they have discovery potential, and in the absence of a signal they allow for constraints to be placed, subject only to the uncertainty in the remnant black hole properties as measured from the merger GW signal. Simple estimates using matched filter signal-to-noise ratios (SNRs) suggest that these types of searches are \textit{in principle} possible for both scalar and vector boson clouds using ground-based detectors~\cite{Chan2022}. In Ref.~\cite{Isi2019}, it was demonstrated, using a CW search method (a hidden Markov model), that follow-up searches for scalar boson clouds around remnant black holes may plausibly be conducted only in the next-generation era of detectors. Vector boson clouds, on the other hand, grow on much faster timescales and radiate at higher power compared to their scalar counterparts, resulting in significantly stronger GW emissions~\cite{East2017, Baryakhtar2017, East2018, Siemonsen2020}; however, they exhibit relatively fast frequency evolution, rendering the detection of these signals with traditional CW search methods challenging~\cite{Isi2019}.

In this paper, we propose a new method to search for long-duration, quasi-continuous GWs produced by ultralight {\em vector} boson clouds. We start by giving an overview of black hole superradiance as it relates to GW science. Making use of the latest waveform model, we detail the boson signal morphology and the parameter space of a typical search and present estimated horizon distances for current and next-generation detectors. We discuss the numerous challenges unique to vector boson searches and propose detailed guidelines for a directed search that is capable of handling these challenges. In addition, we discuss potential target sources with an emphasis on remnant black holes from compact binary merger events observed by the ground-based GW detector network.

The structure of the paper is as follows.
In Sec.~\ref{sec:boson_clouds}, we review the concept of black hole superradiance, the evolution of the boson cloud, and different GW emission mechanisms. We detail the signal waveform and parameter space and the typical duration of a vector boson signal. In Sec.~\ref{sec:directed_searches}, we describe the search method and present guidelines for selecting a configuration and running a directed search for vector boson signals. We obtain horizon distances through a series of simulations using the numerical waveforms in Sec.~\ref{sec:sensitivity}. In Sec.~\ref{sec:sources}, we discuss promising target black holes and the required sky grid spacing.
We conclude in Sec.~\ref{sec:conclusions}. Note, we employ $c=1$ units throughout.

\section{Vector boson clouds}
\label{sec:boson_clouds}

The prospect of probing new physics beyond the Standard Model using the superradiance mechanism has motivated significant progress in understanding the relevant processes~\cite{Brito2020}. In the case of vector boson clouds, a combination of analytic~\cite{Rosa:2011my,Pani2012,Baryakhtar2017,Frolov2018,Baumann:2019eav} and numerical~\cite{East2017,East2017_2,East2018,Dolan2018,Siemonsen2020,Cardoso2018} methods have been applied to predict observational signatures of the presence of this mechanism and the resulting clouds. In the following, we provide a brief overview of the black hole superradiance process in Sec.~\ref{sec:superradiance}, discuss the quasi-monochromatic gravitational radiation properties in Sec.~\ref{sec:emission_&_timescales}, describe the parameterization of the GWs in Sec.~\ref{sec:GW_signals}, and discuss the parameter space of the black hole-boson system and the corresponding GW signal characteristics in Sec.~\ref{sec:param_space}. We mainly focus on the case of vector boson clouds in this study.

\subsection{Superradiant clouds}
\label{sec:superradiance}

In this work, we are primarily interested in following up binary black hole merger events detected with a ground-based GW detector network. The properties of the merger remnant determine the subsequent growth and evolution of the superradiant cloud. After the merger, the cloud begins extracting energy and angular momentum from the remnant, growing exponentially with $e$-folding timescale $\tau_{\rm inst}$. This growth phase saturates after roughly $\sim\mathcal{O}(90)\tau_{\rm inst}$ (for stellar-mass black holes) by spinning down the black hole and emitting gravitational radiation. Provided an ultralight boson of the right mass exists, we expect a detectable GW signal to emerge from the sky-position of the binary merger after roughly $\sim\mathcal{O}(90)\tau_{\rm inst}$. The boson cloud dissipates through GW radiation on a timescale $\tau_{\rm GW}\gg \tau_{\rm inst}$, resulting in a frequency drift (spin up) 
of the nearly monochromatic signal. In this section, we briefly summarize the timescales and frequencies relevant to the growth and saturation of the superradiant cloud and return to a characterization of the emitted GW signal in Secs.~\ref{sec:emission_&_timescales} and~\ref{sec:GW_signals}.

Superradiance around a spinning black hole of mass $M$ and dimensionless spin $\chi$ may be triggered by a collection of ultralight bosons.\footnote{We focus here entirely on the case of ultralight vector bosons. See, e.g., Ref.~\cite{Siemonsen2022} for a comparison of the properties of superradiant scalar and vector boson clouds.} Let this cloud be characterized by an azimuthal mode number $m$ and mode frequency $\omega$. Then a bosonic perturbation extracts rotational energy from the black hole through the superradiance process if the condition
\begin{align}
    0<\omega<m\Omega_H
    \label{eqn:sup_condition}
\end{align}
is satisfied. Here, $\Omega_H=\chi/(2r_+)$ is the black hole horizon frequency, with $r_+=r_g (1 + \sqrt{1 - \chi^2})$ and $r_g=GM$. Due to the vector boson's mass $m_V$, it can form states that are gravitationally bound to the black hole and are thus continuously amplified; this is known as superradiant instability. In order to quantify the relevant timescales and frequencies associated with this process, it is instructive to define the dimensionless ``gravitational fine-structure" constant
\begin{equation}
    \alpha \equiv \frac{r_g}{\lambdabar} = GM \frac{m_V}{\hbar},
    \label{eqn:fine_structure_constant}
\end{equation}
which compares the size of the black hole $\sim r_g$ to the reduced Compton wavelength of the ultralight boson $\lambdabar=\hbar/m_V$. Therefore, $\alpha$ naturally divides the parameter space into a non-relativistic regime, $\alpha\ll 1$, where the superradiant cloud is much larger than the black hole, and a relativistic regime, $\alpha\sim 1$, where the black hole and the cloud are roughly the same size. 

In the non-relativistic regime, the gravitational influence of the black hole follows a simple inverse-radius potential. Hence, the superradiant cloud growing around the black hole is in a hydrogen-like gravitationally bound state. Cloud states are characterized by $m \geq 1$, radial node number $\hat{n}\geq 0$, and polarization $S\in\{-1,0,1\}$. In this limit, the frequency of each state is given by~\cite{Rosa:2011my,Baryakhtar2017,Dolan2018,Baumann:2019eav}
\begin{equation}
    \omega = \frac{m_V}{\hbar} \left(1-\frac{1}{2} \frac{\alpha^2}{n^2}+G_{mS}\right),
    \label{eqn:energy_levels}
\end{equation}
where $n = m+\hat{n}+S+1$. The coefficients $G_{mS}\sim\mathcal{O}(\alpha^4)$ encode the higher-order corrections in the relativistic $\alpha\sim 1$ regime. From Eq.~\eqref{eqn:energy_levels} we see that the superradiant cloud oscillates with frequency $\omega \approx m_V/\hbar$ (i.e., roughly set by the vector boson mass) around the central black hole. The exponential growth of the superradiant cloud is most efficient when the Compton wavelength of the bosonic particle is comparable to the size of the black hole, $\alpha\sim\mathcal{O}(1)$. The growth rates $\Gamma_{\rm inst}$ (and associated timescales $\tau_{\rm inst} \equiv \Gamma^{-1}_{\rm inst}$) for the vector field
in the small-$\alpha$-limit are~\cite{Baryakhtar2017}
\begin{equation}
    \Gamma_{\rm inst} = \alpha^{4m+2S+5}2r_+(m\Omega_H-\omega)C_{mS}.
    \label{eqn:growth_rate}
\end{equation}
The coefficients $C_{mS}>0$ depend on the black hole spin $\chi$~\cite{Baumann:2019eav}, and in the relativistic regime also on $\alpha$~\cite{Siemonsen2020}. In the non-relativistic limit, the growth rates are highly suppressed by large powers of $\alpha$; states with small azimuthal index $|m|\geq 1$, zero radial nodes ($\hat{n}=0$), and $S=-1$ grow the fastest. In this work, we primarily focus on $m=1$ unstable modes. 
For $\omega>m\Omega_H$, the growth rates turn negative, i.e., $\Gamma_{\rm inst}<0$, indicating the exponential decay of this cloud state. Therefore, the superradiance process is most efficient, i.e., $\Gamma_{\rm inst}$ is largest, for $\alpha\lesssim m/2$. 

Both the frequency $\omega$ and growth rates $\Gamma_{\rm inst}$ are computed for superradiant vector clouds in the non-relativistic~\cite{Baryakhtar2017,Baumann:2019eav} (see also Ref.~\cite{Dolan2007}) and relativistic~\cite{East2017,Dolan2018,Cardoso2018,Siemonsen2020} regimes using analytic and numerical methods, respectively. In this work, we utilize the waveform model \texttt{SuperRad}~\cite{Siemonsen2022}, which interpolates between the analytic and numerical results in their different regimes of validity, providing the most accurate estimates for $\omega$ and $\Gamma_{\rm inst}$ which remain valid across the entire relevant parameter space. To provide some intuition, in the non-relativistic limit $\alpha \ll 1$, the instability timescale of the fastest growing mode $(m,\hat{n},S)=(1,0,-1)$ around a black hole of dimensionless spin $\chi$ is roughly given by
\begin{equation}
     \tau_{\rm inst} \approx 2~{\rm mins} \left(\frac{M}{10M_{\odot}}\right) \left(\frac{0.1}{\alpha}\right)^7 \frac{1}{\chi}.
     \label{eqn:growth_timescale}
\end{equation}
We return to the frequency $\omega$ for typical parameters in the next section.

For a given superradiant energy level, as long as Eq.~\eqref{eqn:sup_condition} is satisfied, the occupation number of the vector cloud continues to grow as energy and angular momentum are extracted from the black hole. However, the system eventually reaches the point at which the black hole has lost sufficient energy and angular momentum, i.e., $\Omega_H$ has decreased such that Eq.~\eqref{eqn:sup_condition} becomes asymptotically saturated: $m\Omega_H\rightarrow\omega$. At this stage, the system has reached a quasi-equilibrium state between the black hole and boson cloud~\cite{East2018}. In the absence of any additional processes (e.g., accretion), the saturated mass of the boson cloud $M_c$ is simply the difference between the initial and final black hole masses, $M_i$ and $M_f$, respectively. When only the $m=1$ energy level is populated, the cloud mass is approximately~\cite{Brito2017_2}
\begin{equation}
    M_c = M_i - M_f \approx 0.01\left(\frac{\alpha}{0.1}\right)\left(\frac{\chi_i-\chi_f}{0.1}\right)M_i,
    \label{eqn:cloud_mass_final}
\end{equation}
where $\chi_i$ and $\chi_f$ are the initial and final dimensionless spins of the black hole, respectively. We assumed here that initially the system satisfies $\alpha_i\ll 1$. In this same non-relativistic limit, the spun-down black hole has a final spin of approximately
\begin{equation}
    \chi_f = \frac{4 \alpha_f m}{4 \alpha_f^2 + m^2} < \chi_i,
    \label{eqn:BH_spin_final}
\end{equation}
where $\alpha_f = GM_f m_V/\hbar$. Equations \eqref{eqn:cloud_mass_final} and \eqref{eqn:BH_spin_final} are obtained assuming a linear and adiabatic evolution of the black hole-boson cloud system, as well as the saturation of the superradiance condition $\omega=m\Omega_H$. A more accurate prediction for $M_f$ and $\chi_f$ can be obtained by numerically solving a set of four ordinary differential equations describing the linear evolution of the black hole and boson cloud masses and angular momenta, and the corresponding gravitationally emitted energy and angular momentum~\cite{Brito2015}. While the waveform model \texttt{SuperRad} is able to provide more accurate estimates, in the remainder of this work we assume the system saturates at $\omega=m\Omega_H$.\footnote{Spot checks have been carried out using the more accurate evolution estimates, and no difference is found in the search results.} The final black hole and cloud parameters obtained assuming $\omega=m\Omega_H$, and using the aforementioned linear adiabatic time-domain evolution, differ only at the percent level (see Ref.~\cite{Siemonsen2022}), and hence do not affect any of the conclusions drawn in this work. It has been shown using fully nonlinear numerical relativity techniques that, due to superradiance alone, the cloud can extract up to 10\% of the black hole's total mass~\cite{East2017_2}. Once the growth saturates, GW emission stemming from the dissipation of the oscillating cloud becomes the dominant evolution mechanism. This occurs over timescales much longer than those of the cloud growth.

\subsection{GW emission}
\label{sec:emission_&_timescales}

In the context of isolated boson clouds, there are several mechanisms that may produce gravitational radiation: emission from boson annihilation in a single cloud state, the transition of bosons in the cloud between energy levels, and the collapse of the boson cloud under its own self-interactions, i.e., a ``bosenova.” Cloud transitions, or a beating between modes oscillating at different frequencies, may only produce significant gravitational radiation if comparable occupation numbers can be found in more than one energy level. In the case of vector clouds, these transitions can occur even for young black holes, resulting in quasi-periodic GW signals (whereas transitions in scalar clouds occur primarily around old black holes~\cite{Arvanitaki2015, Arvanitaki2017}). Unfortunately, because these transitions occur on timescales shorter than that of the typical GW emission of a single vector cloud level, the observational window for transition signals is small~\cite{Siemonsen2020}. If the massive vector boson obtains its mass through a Higgs mechanism, depending on the relevant coupling constants, the cloud can reach a large enough occupation number to backreact on the Higgs-like field and lead to the formation of string vortices. This drives a stringy bosenova~\cite{East:2022ppo, East:2022rsi} where the cloud is strongly disrupted, resulting in recurring burst-like GW emission.\footnote{In the case of axions, these explosive phenomena due to self-interactions are unlikely to happen~\cite{Yoshino2012, Yoshino2015_2, Baryakhtar:2020gao, Omiya2022}.} However, these explosive phenomena are not yet well-modelled. In this work we focus solely on quasi-monochromatic GW emissions from the dissipation of a single cloud state assuming only the gravitational coupling. This picture may change, however, if the vector boson couples sufficiently strongly to the Standard Model, or to an extended dark sector~\cite{Fukuda:2019ewf,Caputo:2021efm,Cannizzaro:2022xyw,Siemonsen:2022ivj}.

As discussed in the previous section, once the superradiant growth of the cloud terminates with the saturation of the condition in Eq.~\eqref{eqn:sup_condition}, the cloud of mass $M_c$ oscillates around the black hole with angular frequency $\omega$. The frequency of the emitted quasi-monochromatic GWs is then set by $\omega$, defined in Eq.~\eqref{eqn:energy_levels}, in the source frame as twice the mode frequency:
\begin{equation}
    f_{\rm GW} = \omega / \pi.
    \label{eqn:GW_angular_frequency}
\end{equation}
For typical cloud parameters in the non-relativistic limit, and considering only the most unstable $m=1$ mode, the GW signal frequency is roughly [Eqs.~\eqref{eqn:fine_structure_constant}--\eqref{eqn:energy_levels}]
\begin{equation}
    f_{\rm GW} \approx 645~{\rm Hz} \left(\frac{10\ M_{\odot}}{M}\right) \left(\frac{\alpha}{0.1}\right).
    \label{eqn:GWfrequency}
\end{equation}
Stellar-mass black holes can therefore support boson clouds with emission frequencies that lie within the most sensitive band of ground-based GW detectors. (We return to the frequency evolution later in this section.)

The GW amplitude is obtained on a Kerr black hole background using the Teukolsky formalism~\cite{Teukolsky1973} (or limits thereof). The non-relativistic estimate obtained in Ref.~\cite{Baryakhtar2017} is refined and extended to the relativistic regime in Ref.~\cite{Siemonsen2020} and is consistent with a fully nonlinear treatment of the problem~\cite{East2017,East2018}. Combining these results, in the non-relativistic regime and for the $m=1$ cloud state, the characteristic amplitude is roughly
\begin{equation}
    h_0 \approx 3 \times 10^{-26} \left(\frac{M}{10\ M_{\odot}}\right) \left(\frac{\alpha}{0.1}\right)^{5} \left(\frac{0.1\ \rm Gpc}{d}\right) \left(\frac{\chi_i - \chi_f}{0.1}\right).
    \label{eqn:characteristic_amplitude_approx}
\end{equation}
Here, we define $d$ as the luminosity distance, $h_0=(10 \dot{E}_{\rm GW})^{1/2}/(2\pi f_{\rm GW}d)$ as the characteristic GW strain, and $\dot{E}_{\rm GW}$ as the total GW energy flux. We can approximate the power radiated by the fastest-growing cloud state as
\begin{equation}
    \dot{E}_{\rm GW} \approx 6 \times 10^{46} ~\mathrm{erg/s} \left(\frac{\alpha}{0.1}\right)^{12} \left(\frac{\chi_i-\chi_f}{0.1}\right)^2
    \label{eqn:GW_power_approx}
\end{equation}
in the non-relativistic regime. Note that the radiated power from vector clouds, which are the focus of this paper, is orders of magnitude larger than that from scalar clouds. This also implies, however, that the signals from vector boson clouds last for timescales orders of magnitude shorter than those from scalar clouds. (We return to this point and its implications for the search methods below.) 

As the cloud dissipates energy to GWs, the signal amplitude, starting at its peak when the cloud is at its maximum size [given by Eq.~\eqref{eqn:characteristic_amplitude_approx}], will decrease over time. This occurs over a timescale on the order of the signal duration. During the emission process, the cloud's mass $M_c$ decreases following the relationship
\begin{align}
    M_c(t)=\frac{M_c^{\rm sat}}{1+(t-t_{\rm sat})/\tau_{\rm GW}}, & & \tau_{\rm GW} \equiv \frac{M_c^{\rm sat}}{\dot{E}_{\rm GW}^{\rm sat}},
    \label{eqn:emission_timescale}
\end{align}
where $t_{\rm sat}$ is the saturation time of the superradiant growth, and $M_c^{\rm sat}$ and $\dot{E}_{\rm GW}^{\rm sat}$ are the quantities at $t_{\rm sat}$, and we define the characteristic GW emission timescale $\tau_{\rm GW}$. We consider this to be the typical duration of the signal, since at a time $\tau_{\rm GW}$ after the saturation, the signal amplitude has halved. Recalling the approximations made in Eq.~\eqref{eqn:cloud_mass_final} and Eq.~\eqref{eqn:GW_power_approx}, we can rewrite this timescale in the dominant vector energy level and in the non-relativistic limit as
\begin{equation}
    \tau_{\rm GW} \approx 33 ~\mathrm{days} \left(\frac{M}{10M_{\odot}}\right) \left(\frac{0.1}{\alpha}\right)^{11} \left(\frac{0.1}{\chi_i-\chi_f}\right).
    \label{eqn:emission_timescale_approx}
\end{equation}
The GW emission timescale is typically orders of magnitude longer than the cloud growth timescale [compare Eq.~\eqref{eqn:growth_timescale} to Eq.~\eqref{eqn:emission_timescale_approx}], which allows us to treat the two as distinct stages in the system's evolution. 

Lastly, the evolution of the total mass of the cloud [$M_c(t)$ in Eq.~\eqref{eqn:emission_timescale}] implies an increase in the GW frequency $f_{\rm GW}$ on the timescale $\tau_{\rm GW}$. The smaller $M_c$ is, the weaker the gravitational redshift of the emitted GW becomes, and hence the higher the frequency. The resulting frequency drift can, to leading order, be approximated by the change of the Newtonian potential sourced by the presence of the superradiant cloud~\cite{Baryakhtar2017, Isi2019, Siemonsen2020, Baryakhtar:2020gao}.\footnote{Note a missing factor of 2 in the frequency drift expressions of Refs.~\cite{Baryakhtar2017, Isi2019} is pointed out in Ref.~\cite{Baryakhtar:2020gao} and included in the relevant vector cloud expressions in Ref.~\cite{Siemonsen:2022ivj}.} The first time derivative of the frequency, $\partial_t^{(1)}f_{\rm GW}\equiv\dot{f}_{\rm GW}$, can be obtained from the rate of change of $M_c$ at the saturation point and is given by~\cite{Siemonsen:2022ivj}
\begin{align}
\begin{aligned}
    \dot{f}_{\rm GW} = & \ \frac{5 \alpha^3 G}{8\pi r_g^2} \dot{E}_{\rm GW}\\
    \approx & \ 10^{-8} ~\mathrm{Hz/s} \left(\frac{10\ M_{\odot}}{M}\right)^2 \left(\frac{\alpha}{0.1}\right)^{15} \left(\frac{\chi_i-\chi_f}{0.1}\right)^2
    \label{eqn:fdot}
\end{aligned}
\end{align}
in the $\alpha \ll 1$ limit. The complete frequency evolution with all time derivatives $\partial_t^{(n)} f_{\rm GW}$ can be determined directly from Eq.~\eqref{eqn:emission_timescale}.

In this section, we focused on providing intuition for and rough scalings of all relevant observables. In practice, however, we utilize \texttt{SuperRad}~\cite{Siemonsen2022} to accurately determine the GW amplitude, frequency evolution, and involved timescales across the entire parameter space for a set of initial source parameters. 

\subsection{GW signal parameterization}
\label{sec:GW_signals}

In the previous section, we qualitatively introduced all relevant GW properties. In the following section, we discuss the precise parameterization of the quasi-monochromatic gravitational radiation used in the remainder of this work. We focus on the source frame quantities, and we assume that a single cloud level is dominating the GW signal and that the superradiance condition is saturated. 

The GW strain signal in a detector $I$ is written as a sum over two polarizations:
\begin{equation}
    h^I(t) = F_{+}^I(t) h_+(t)+F_{\times}^I(t) h_\times(t),
    \label{eqn:strain_signal}
\end{equation}
where $F_{+}^I$ and $F_{\times}^I$ are the antenna response (or beam pattern) functions of detector $I$ to GW signals with plus ($+$) and cross ($\times$) polarizations, respectively. (For explicit expressions, see e.g., Appendix B in Ref.~\cite{Anderson2001}.) These are periodic functions that depend on the relative location of the detector and source, typically parameterized by right ascension (RA) and declination (Dec), and the polarization angle. The strain amplitudes $h_{+,\times}(t)$ are set by the GW phase and amplitude and the inclination angle ($\iota$) between the rotational axis of the black hole and the line of sight.

We expand the GW polarization waveforms $h_{+,\times}$ in terms of a series of spin-weighted spherical harmonics ${}_s Y_{\tilde{\ell} \tilde{m}}(\iota,\varphi)={}_s S_{\tilde{\ell} \tilde{m}}(\iota) e^{-i\tilde{m}\varphi}$ of spin-weight $s=-2$, where $\varphi$ is the azimuthal coordinate in the source frame. With the luminosity distance $d$, the polarization waveform is generally (see, e.g., Ref.~\cite{Siemonsen:2022ivj})
\begin{align}
\begin{aligned}
    h_+= & \ \frac{1}{d}\sum_{\tilde{\ell}\geq \tilde{m}}|h^{\tilde{\ell} \tilde{m}}|({}_{-2}S_{\tilde{\ell} \tilde{m}}+(-1)^{\tilde{\ell}} {}_{-2}S_{\tilde{\ell} -\tilde{m}})\\ 
    & \qquad \times\cos(\Phi(t)+\tilde{m}\varphi+\phi_{\tilde{\ell} \tilde{m}}),
    \label{eqn:h_plus}
\end{aligned}
\\
\begin{aligned}
    h_\times= & \ \frac{1}{d}\sum_{\tilde{\ell}\geq \tilde{m}}|h^{\tilde{\ell} \tilde{m}}|({}_{-2}S_{\tilde{\ell} \tilde{m}}-(-1)^{\tilde{\ell}} {}_{-2}S_{\tilde{\ell} -\tilde{m}})\\
    & \qquad \times\sin(\Phi(t)+\tilde{m}\varphi+\phi_{\tilde{\ell} \tilde{m}}).
    \label{eqn:h_cross}
\end{aligned}
\end{align}
Here, $\Phi(t)$ is the GW phase, $h^{\tilde{\ell} \tilde{m}}$ is the GW mode amplitude, and $\phi_{\tilde{\ell} \tilde{m}}$ is the phase difference between the $(\tilde{\ell},\tilde{m})$-modes. (By construction, we choose $\phi_{22}=0$, while all other $\phi_{\ell m}$ may be non-vanishing.) In the non-relativistic limit ($\alpha\ll 1$), the GW signal from an $m$ cloud state is dominated by the $\tilde{\ell}=\tilde{m}=2m$ contribution; for $m=1$ vector cloud solutions around the black hole, the GW emission is given entirely by the $(\tilde{\ell},\tilde{m})=(2,2)$ component. In the relativistic regime ($\alpha\sim 1$), this mode is subdominant to the $(\tilde{\ell},\tilde{m})=(3,2)$ contribution. Similar behavior can be observed for higher-$m$ cloud states~\cite{Siemonsen2020}.

The frequency evolution of the quasi-monochromatic GW signal is encoded in the GW frequency
\begin{align}
    f_{\rm GW}(t)=f_0+\frac{\delta f}{1+(t-t_{\rm sat})/\tau_{\rm GW}},
\label{eqn:fgw}
\end{align}
where $f_0$ is the asymptotic GW frequency at late times ($t \gg \tau_{\rm GW}$),
and $\delta f$\footnote{For an explicit form of $\delta f=\Delta \omega/\pi$, see Eq. (22) in Ref.~\cite{Siemonsen2022}.} characterizes the shift of $f_{\rm GW}$ away from $f_0$ due to the self-gravity of the superradiant cloud, with $|f_{\rm GW}(t)-f_0| \propto M_c$. In all cases considered here, $f_0\gg |\delta f|$. The frequency derivative $\dot{f}_{\rm GW}$, given in Eq.~\eqref{eqn:fdot}, follows directly from Eq.~\eqref{eqn:fgw}. The GW phase $\Phi(t)$ is simply the integral of $f_{\rm GW}$ starting from the saturation time $t_{\rm sat}$ up to time $t$, added to an initial phase-offset $\Phi_0$:
\begin{align}
    \Phi(t)=\Phi_0+2\pi \int_{t_{\rm sat}}^t dt' f_{\rm GW}(t').
\end{align}
The frequency and phase evolution in the source frame are modified at cosmological distances by appropriate redshift factors. We discuss the details of this in Sec.~\ref{sec:horizon_distance}.

Finally, here and in the following, we use $\hpeak$ and $\fpeak$ to denote the signal strain amplitude and frequency, respectively, at the time of cloud saturation, i.e., $\hpeak = h_0(t_{\rm sat})$ and $\fpeak = f_{\rm GW}(t_{\rm sat})$.

\subsection{System parameter space and GW emission characteristics}
\label{sec:param_space}

In the following, we expand on the presentation in Secs.~\ref{sec:superradiance} and~\ref{sec:emission_&_timescales} and analyze the relevant parameter space in more detail, focusing entirely on the most unstable $m=1$ vector boson cloud state. The expected GW signal parameters depend on the intrinsic parameters of the system---the initial mass and spin of the black hole and the boson mass. For a given black hole, there exists a broad range of boson masses that satisfy the superradiance condition. Since in a real search the vector boson mass is unknown, a single black hole with known intrinsic parameters allows us to probe a range of boson masses and requires us to consider a corresponding range of GW frequencies, frequency evolution rates, emission timescales, and signal amplitudes.

\begin{figure}[hbt!]
	\centering
	\includegraphics[scale=.49]{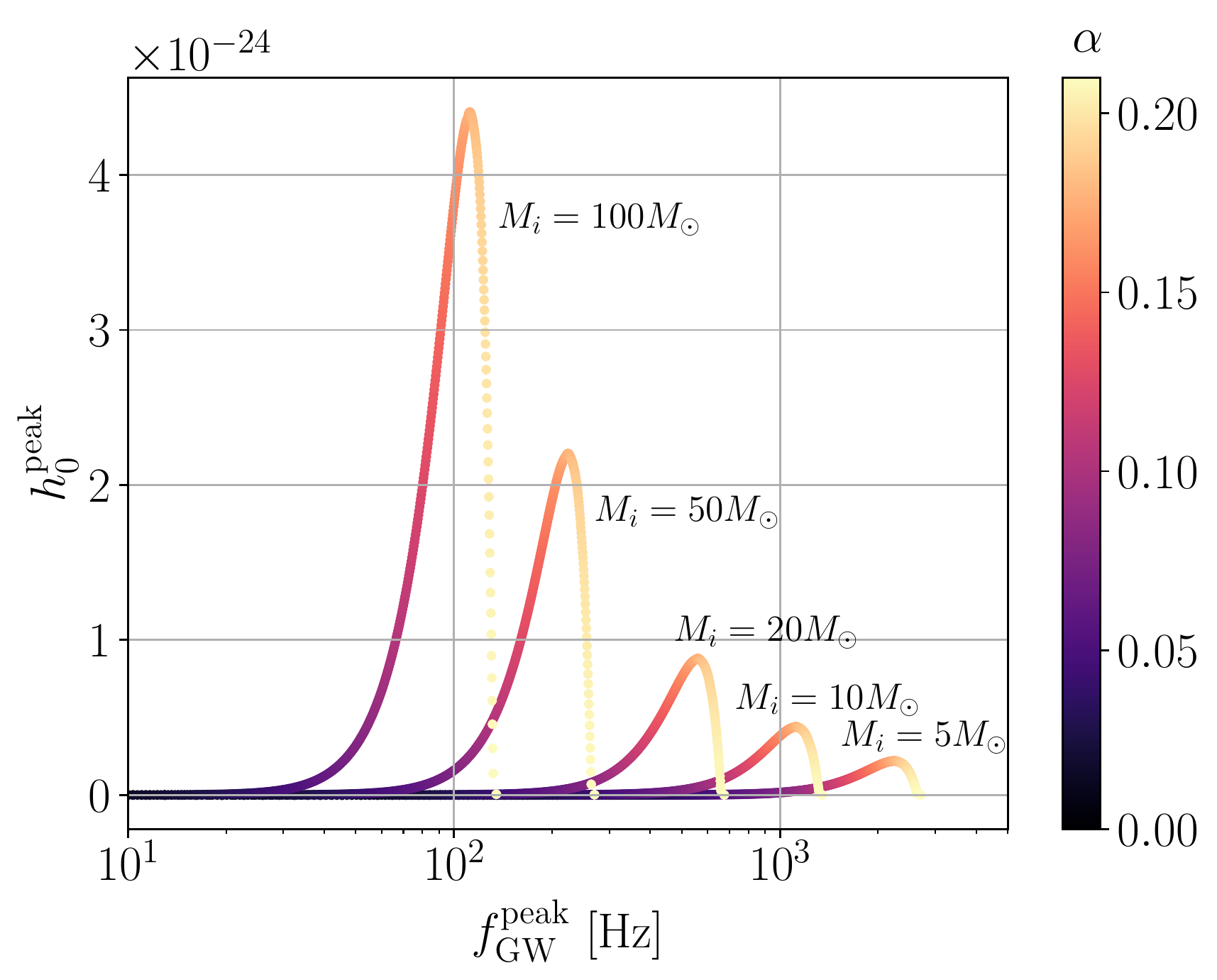}
	\caption{Strain amplitude at the cloud's saturation $\hpeak$ as a function of $\fpeak$ for five different black holes with $M_i$ = 5, 10, 20, 50, and 100~$M_\odot$ for $\chi_i = 0.7$ and $d = 100$~Mpc. The color corresponds to $\alpha$. For all five black holes, the optimally matched scenario (i.e., maximum $\hpeak$) occurs at $\alpha_{\rm opt} = 0.176$.}
	\label{fig:frequency_v_h0_v_alpha}
\end{figure}

Figure~\ref{fig:frequency_v_h0_v_alpha} shows the characteristic GW strain at saturation $\hpeak$ as a function of emission frequency $\fpeak$ for a vector cloud (in the most unstable $m=1$ cloud state) around five different black holes with masses from $5~M_{\odot}$ to $100~M_{\odot}$ (with initial spin $\chi_i = 0.7$ and luminosity distance $d = 100$~Mpc). For convenience, we parameterize how ``well-matched" the boson mass is to the black hole via the gravitational fine-structure constant $\alpha$ [Eq.~\eqref{eqn:fine_structure_constant}]. We define $\alpha_{\rm opt}$ to be the fine-structure constant which maximizes the strain amplitude in the source frame for a fixed initial black hole mass and spin, i.e.,
\begin{align}
    \alpha_{\rm opt} = \arg {\max}_\alpha \hpeak.
    \label{eq:alphaopt}
\end{align}
For each given black hole in Fig.~\ref{fig:frequency_v_h0_v_alpha}, $\alpha = \alpha_{\rm opt}$ only when $\hpeak$ is at its maximum. (This will not in general correspond exactly to the $\alpha$ value that gives the largest horizon distance; see Sec.~\ref{sec:non-optimal}.)
As an example, the signal strain for $M_i = 100~M_{\odot}$, $\alpha_{\rm opt} = 0.176$, and $\cos{\iota} = 1.0$ is estimated to be \mbox{$\hpeak = 4.41 \times 10^{-24}$}. 
For each black hole mass, there is a range of possible $\alpha$ values that allow for superradiance, and accordingly, a range of possible signal parameters. As shown in Secs.~\ref{sec:superradiance} and~\ref{sec:emission_&_timescales} and demonstrated in the figure, more massive black holes emit louder GW signals at lower frequencies.

\begin{figure}[hbt!]
	\centering
	\includegraphics[scale=.44]{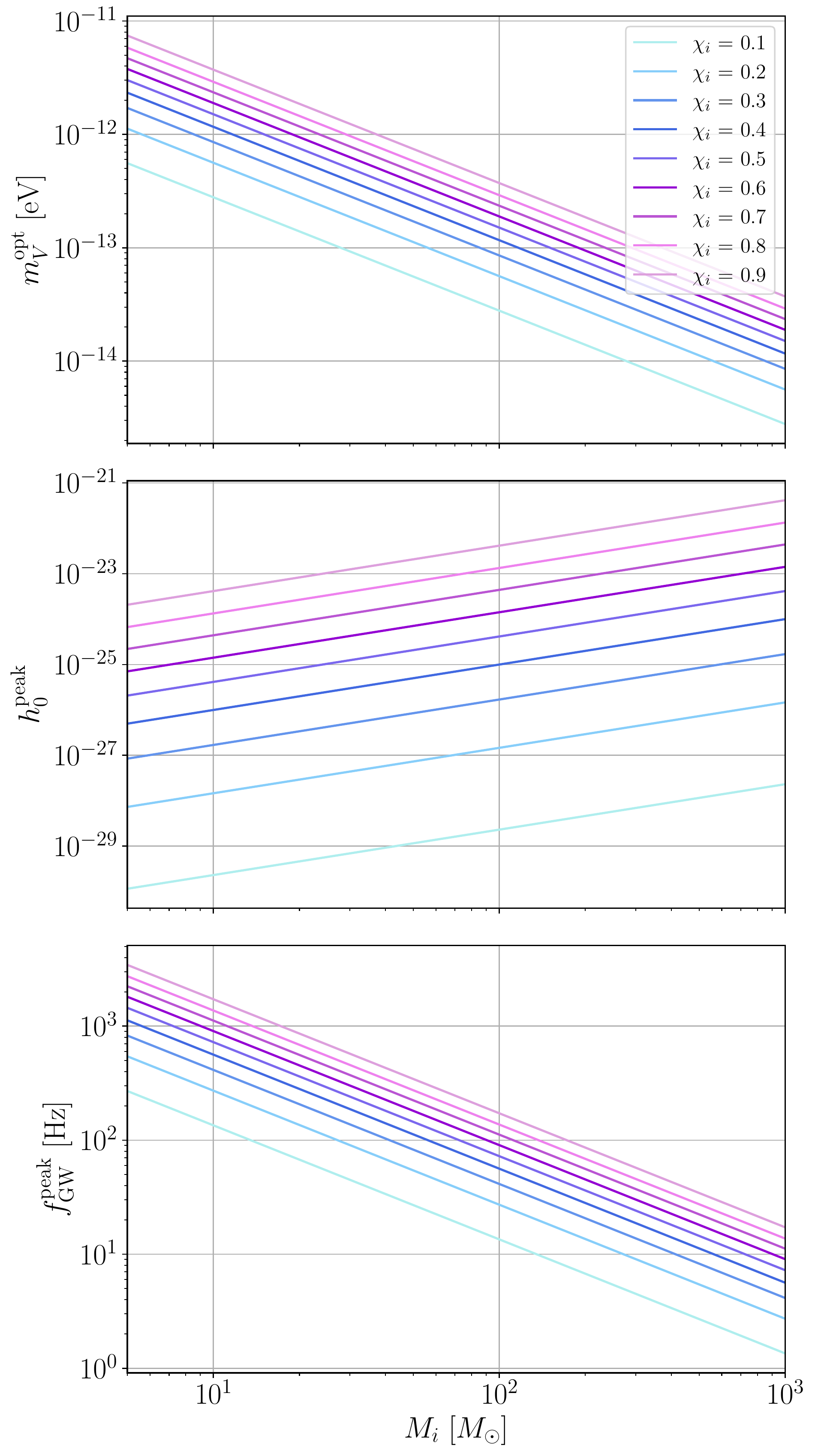}
	\caption{Boson mass $m_V^{\rm opt}=\alpha_{\rm opt}\hbar/r_g$ in units of eV (top), $\hpeak$ (for $d=100$~Mpc) (middle), and $\fpeak$ (bottom) as a function of initial black hole mass $M_i$. In all three panels, we use the optimally matched $\alpha$ as defined in Eq.~\eqref{eq:alphaopt}. The colored lines in each panel correspond to different values of initial black hole spin.}
	\label{fig:optimal_mu_&_h0_&_frequency}
\end{figure}


\begin{figure}[hbt!]
	\centering
	\includegraphics[scale=.42]{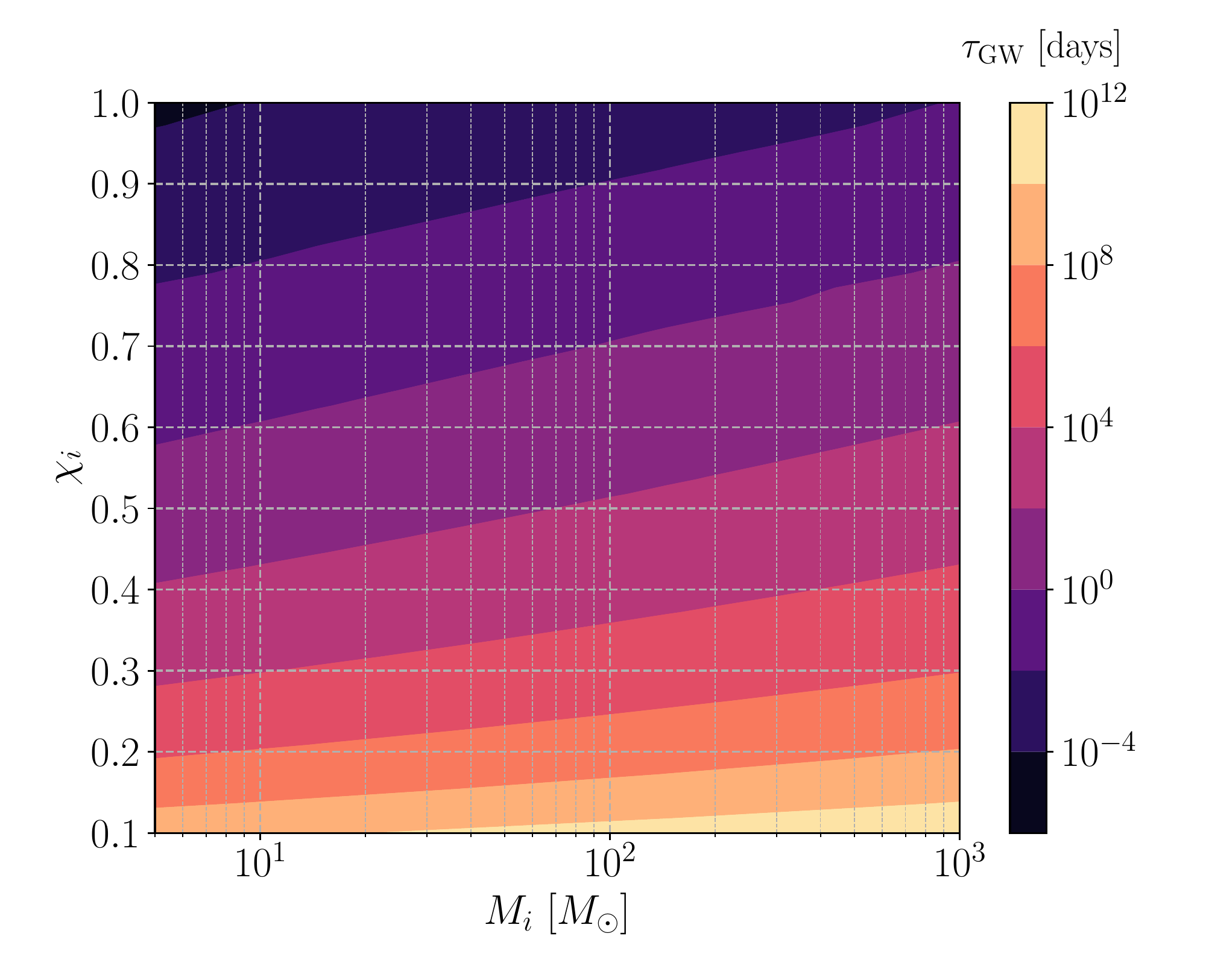}
	\caption{GW emission timescale $\tau_{\rm GW}$ as a function of initial black hole mass $M_i$ and spin $\chi_i$ for $\alpha_{\rm opt}$ as defined in Eq.~\eqref{eq:alphaopt}.}
	\label{fig:tauGW_v_M_v_chi}
\end{figure}

\begin{figure}[hbt!]
	\centering
	\includegraphics[scale=.48]{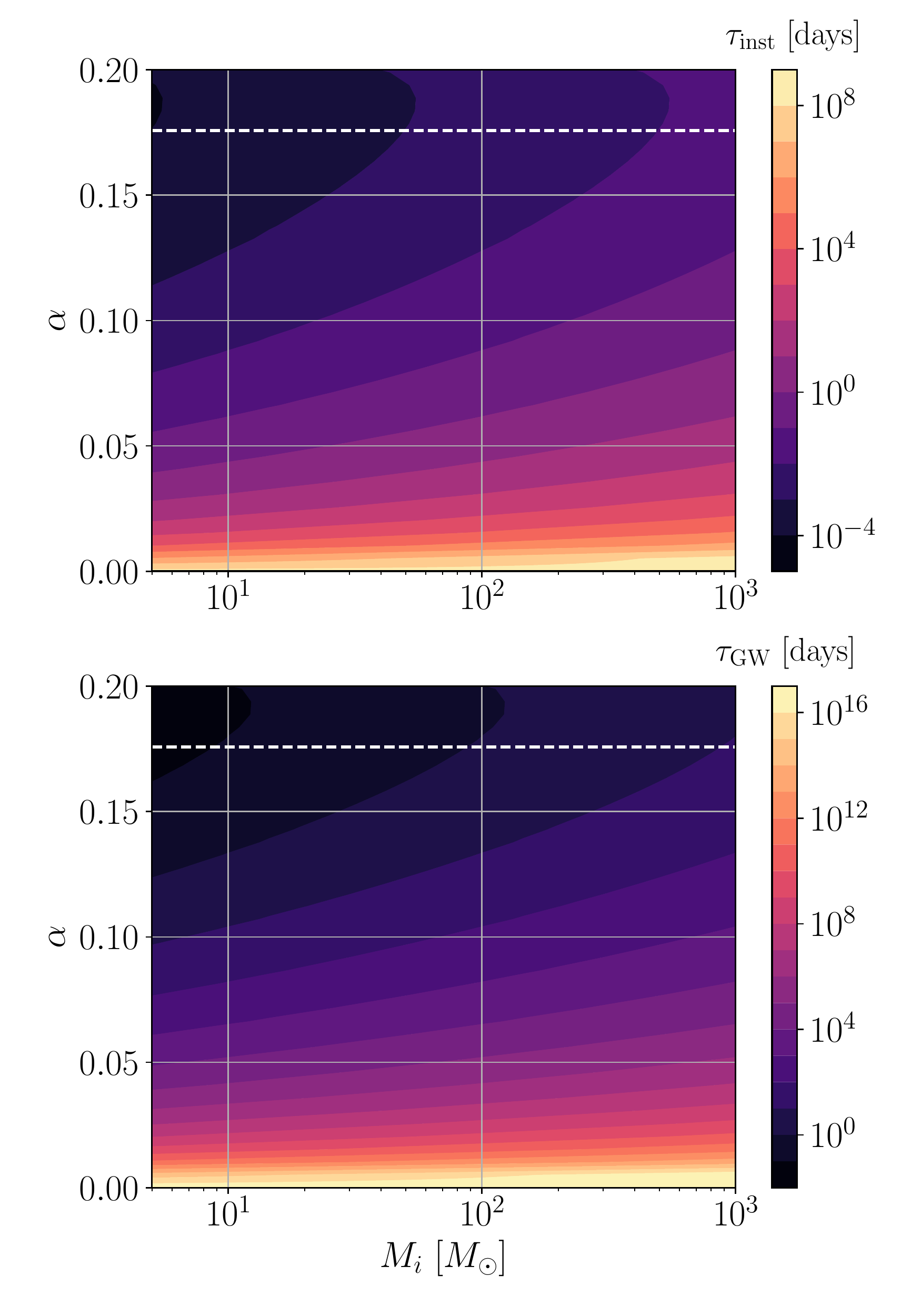}
	\caption{Cloud growth (top) and GW emission timescales (bottom) as a function of initial black hole mass $M_i$ and $\alpha$ with initial black hole spin $\chi_i=0.7$. The dashed white line marks $\alpha=\alpha_{\rm opt} = 0.176$ corresponding to the optimally matched boson mass for each black hole mass, as defined in Eq.~\eqref{eq:alphaopt}.}
	\label{fig:timescales_v_M_v_alpha}
\end{figure}

\begin{figure}[hbt!]
	\centering
	\includegraphics[scale=.42]{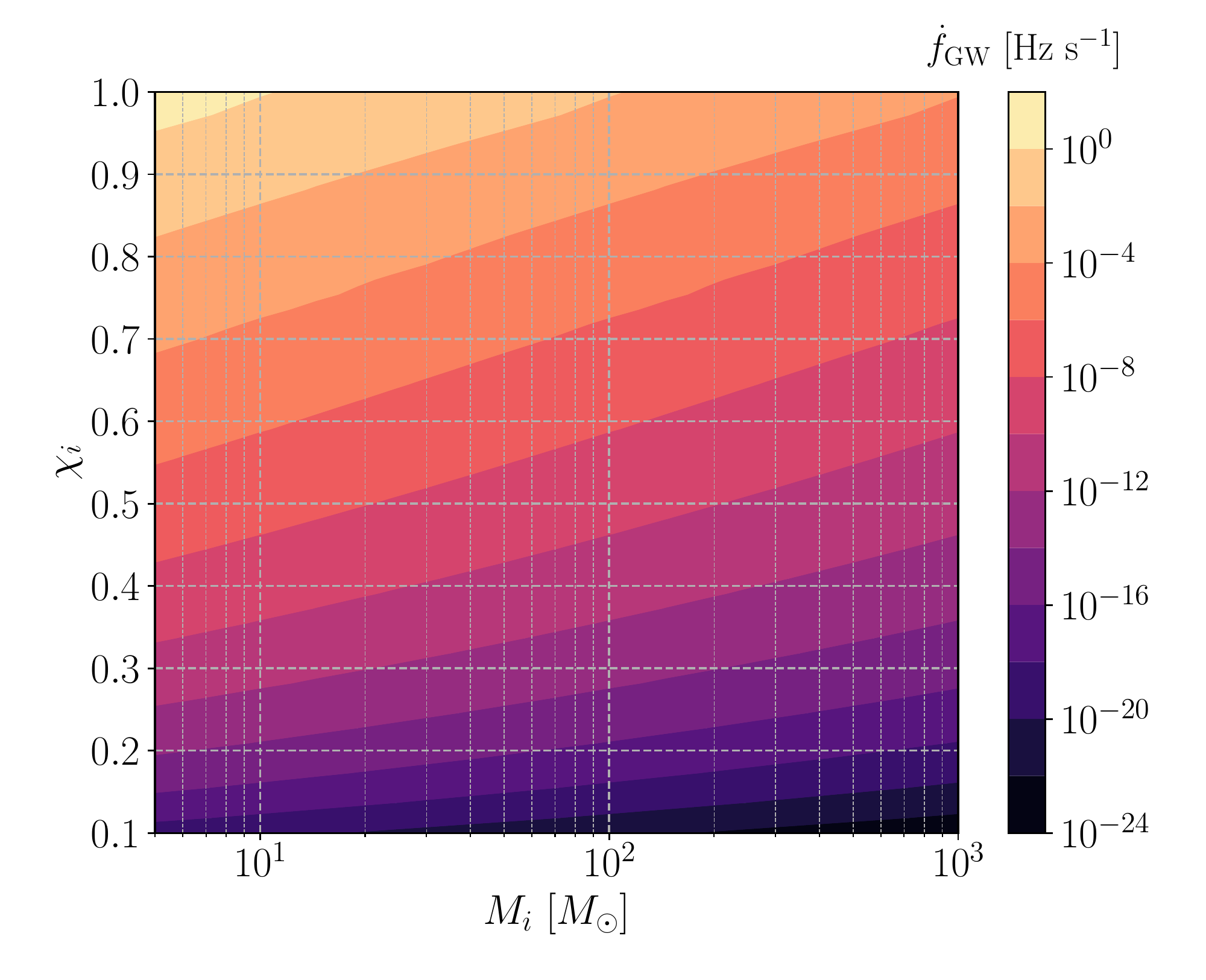}
	\caption{First time derivative of the GW frequency $\dot{f}_{\rm GW}$ (at saturation) as a function of initial black hole mass $M_i$ and spin $\chi_i$, assuming $\alpha = \alpha_{\rm opt}$ as defined in Eq.~\eqref{eq:alphaopt}.}
	\label{fig:fdot_v_M_v_chi}
\end{figure}

Let us consider the impact of the black hole's spin on the emitted GW signal from superradiant vector boson clouds. Figure~\ref{fig:optimal_mu_&_h0_&_frequency} shows the optimally matched boson mass $m_V^{\rm opt}$ as well as $\hpeak$ and $\fpeak$ as a function of initial black hole mass $M_i$ for different values of initial black hole spin $\chi_i$. Generally, as $\chi_i$ increases, $\alpha_{\rm opt}$ (and the corresponding $m_V^{\rm opt}$) increases as well, resulting in an upwards trend of $\hpeak$ and $\fpeak$ [consult also Eqs.~\eqref{eqn:fine_structure_constant}, \eqref{eqn:GWfrequency}, and \eqref{eqn:characteristic_amplitude_approx}], which span a wide range depending on the value of $M_i$.
Similar to what was shown in Fig.~\ref{fig:frequency_v_h0_v_alpha}, heavier black holes together with lighter vector bosons harbor clouds that emit louder GW signals at lower frequencies. 
As dictated by Eq.~\eqref{eqn:cloud_mass_final}, heavier black holes with higher initial spins are able to support heavier boson clouds, resulting in gravitational radiation with higher signal amplitudes.

We now turn to the two timescales characterizing the evolution of the superradiant cloud. In Secs.~\ref{sec:superradiance} and \ref{sec:emission_&_timescales}, we introduced the cloud growth timescale, $\tau_{\rm inst}$, as well as the GW emission timescale, $\tau_{\rm GW}$, which determine how long the vector boson cloud takes to grow, and how long the subsequent GW emission phase lasts. In Fig.~\ref{fig:tauGW_v_M_v_chi}, we show the GW emission timescale $\tau_{\rm GW}$ for a range of initial black hole masses and spins assuming $\alpha = \alpha_{\rm opt}$. The decay of the GW emission after saturation can occur on timescales as short as $\tau_{\rm GW}\sim$~mins for light and rapidly spinning black holes. Dropping the assumption of Eq.~\eqref{eq:alphaopt}, in Fig.~\ref{fig:timescales_v_M_v_alpha} we compare both timescales for a moderate initial spin of $\chi_i = 0.7$. For a given initial black hole mass, as $\alpha$ deviates from $\alpha_{\rm opt}$ (dashed white line), the instability growth and GW emission timescales increase. Moreover, for any given set of $M_i$ and $\alpha$, $\tau_{\rm inst}$ is orders of magnitude smaller than $\tau_{\rm GW}$ for the most unstable $m=1$ cloud state. 

We now consider targeting a newly born black hole, such as a binary black hole merger remnant. It is important to account for the total growth time $t_{\rm growth}=\tau_{\rm inst}\log(M_c/m_V)/2$ of the vector boson cloud around the black hole. Once the cloud reaches its saturation (at an age of $t_{\rm growth}$), it contains a significant fraction of the black hole’s mass, after which the GW signal is most likely to be detectable. Thus, for a stellar-mass black hole, GW emissions from the newly formed vector boson cloud reach their peak around $t_{\rm growth}=\tau_{\rm inst}\log(0.01\ M_\odot/10^{-13}~{\rm eV})/2\approx 85\tau_{\rm inst}$. The amplitude of the GW signal then drops by half roughly $t_{\rm growth}+\tau_{\rm GW}$ after the birth of the black hole. Therefore, a GW search would begin roughly $t_{\rm growth}$ after the black hole is born and last for $\sim\tau_{\rm GW}$. As shown in Fig.~\ref{fig:timescales_v_M_v_alpha}, the growth timescale for $\alpha \gtrsim 0.05$ lasts only a few days or less, which enables us, in most cases, to follow up a black hole merger remnant within the same observing run in which it was detected. If a potential target is observed towards the end of an observing run or right before a significant commission break, however, these timescales may extend beyond the end of the observing period, where we no longer have data. In that case, higher azimuthal states of the vector boson cloud around a given black hole could be considered in the next observing period, but we leave this to future work.

The GW emission timescale is closely related to the GW signal evolution as demonstrated in Secs.~\ref{sec:emission_&_timescales} and~\ref{sec:GW_signals}. In Fig.~\ref{fig:fdot_v_M_v_chi}, we show the first time derivative of the GW signal frequency $\dot{f}_{\rm GW}$ as a function of $M_i$ and $\chi_i$ for $\alpha_{\rm opt}$ at saturation. While the frequency evolution is not linear (see Sec.~\ref{sec:GW_signals}), $\dot{f}_{\rm GW}$ provides a guide to the evolution timescales of the GW frequency. The frequency derivative spans tens of orders of magnitude across the entire parameter space, with lower-mass and higher-spin black holes yielding the largest $\dot{f}_{\rm GW}$ values (for $\alpha_{\rm opt}$). This implies that a vector boson cloud may emit nearly monochromatic CW signals (for high black hole masses and low spins) or be highly dynamical with $\dot{f}_{\rm GW}\sim \mathcal{O}(1)$~Hz~s$^{-1}$ (for low black hole masses and high spins). As we show in the following sections, the search techniques developed in this paper can track signals with $\dot{f}_{\rm GW}$ up to $\sim 10^{-4}$~Hz~s$^{-1}$, covering most of the parameter space in Fig.~\ref{fig:fdot_v_M_v_chi} other than the top left corner.
As a final note, when considering vector signals with small $\dot{f}_{\rm GW}$ values at $t_{\rm sat}$ (towards the bottom right corner in Fig.~\ref{fig:fdot_v_M_v_chi}), their frequency evolution rates are comparable to signals from scalar clouds (but they correspond to very different black holes). In these cases, signals from vector clouds are still generally higher in amplitude, occur over shorter timescales, and are more easily detectable than scalar signals with comparable $\dot{f}_{\rm GW}$ values.

\section{Directed Searches}
\label{sec:directed_searches}

While there are many observational signatures that would allow us to infer the existence of vector boson clouds formed via black hole superradiance~\cite{Arvanitaki2011, Yoshino2014, Yoshino2015, Arvanitaki2015, Arvanitaki2017, Brito2017, Brito2017_2, Baryakhtar2017, Cardoso2018, Baumann2019, Hannuksela2019, Zhang2019, Dantonio2018}, in this study we choose to focus on direct detection via GW radiation. Although the emission timescales for vector bosons are often shorter than the typical timescales associated with CW searches, much of the parameter space (as we show in Sec.~\ref{sec:param_space}) would still produce signals that are considered long-duration. As such, we need to use search techniques that are capable of tracking signals on timescales shorter than CWs but longer than transients. In particular, we focus on directed searches, in which we target black holes with known (or well-constrained) parameters such as mass, spin, and sky position, as opposed to a blind all-sky search, in which we search for signals from unknown black holes with unknown parameters. One benefit of conducting a directed search is that we will be able to place constraints on the existence of vector bosons without needing to rely on black hole population models, which come with large uncertainties. If we make a detection, we will learn the particle's mass and dynamics through detailed measurements of the signal morphology. If no detection is made, having prior knowledge about the black hole's parameters will allow us to place stringent constraints on the boson mass.

In this section, we introduce a hidden Markov model (HMM)-based search method to track vector boson signals. In Sec.~\ref{sec:search_methods}, we review the general HMM algorithm and describe the new implementation of the algorithm in this study to search for vector boson signals. We investigate the parameter space and corresponding configurations of a typical search in Sec.~\ref{sec:simulations} and describe the simulations using \texttt{SuperRad}.

\subsection{Search method}
\label{sec:search_methods}

We implement a semicoherent search method dedicated for vector boson signals, which are expected to be much shorter (on a timescale of hours to months) than the typical CW signal. 
The method combines a frequency-domain matched filter, $\mathcal{F}$-statistic (widely used in CW searches~\cite{Riles2017, JKS, Cutler2005}), with an efficient HMM search technique to track the signal evolution. The HMM tracking technique has been applied to searches for many types of quasimonochromatic, continuous, or long-transient GW signals~\cite{Suvorova2016, Sun2018, Isi2019, Sun2019}. It is an ideal search strategy for signals from vector boson clouds because it is extremely computationally efficient, allowing us to cover a wide parameter space, including signal frequency, duration, and sky position. Other semicoherent search techniques generally rely on Taylor expansions of the signal phase evolution within the matched filtering and are thus quite model-dependent (e.g., Refs.~\cite{JKS, Dhurandhar2008}). HMM, on the other hand, is an ideal choice in searches where uncertainties may exist in the signal waveforms predicted by theories and numerical calculations because it allows for some uncertainty in the signal morphology.

Factoring in the random noise present in the detector data, HMM is able to find the most probable signal frequency evolution, or ``path," as a function of time~\cite{Suvorova2016, Sun2018}. To accomplish this, the frequency-time plane is divided into a discrete grid of $N_Q$ frequency bins and $N_T$ time steps. The length of each time step and the corresponding width of the bins are chosen carefully, based on prior knowledge of the target signal, to satisfy two criteria: the signal is considered ``monochromatic'' (with the signal power concentrated in one bin) over the course of a single time step, and it can move at most one bin from one discrete time step to the next. In other words, the signal does not evolve too rapidly for HMM to track. 

Over the total observing time $T_{\rm obs}$, we select a coherent time interval, $T_{\rm coh}$ = $T_{\rm obs}/N_T$, such that
\begin{equation}
    \left|\int_t^{t+T_{\rm coh}}dt' \dot{f}_{\rm GW}(t')\right| \leq \Delta f
    \label{eqn:Tcoh_int}
\end{equation}
is always satisfied for $0 < t < T_{\rm obs} - T_{\rm coh}$, 
i.e., the signal does not evolve outside the tracking capabilities of HMM. Here $\Delta f = 1/(2T_{\rm coh})$ is the frequency bin size in the $\mathcal{F}$-statistic output computed over $T_{\rm coh}$, where the $\mathcal{F}$-statistic calculation takes a series of short Fourier transforms (SFTs) of length $T_{\rm SFT}$ $(< T_{\rm coh})$ as input.
Considering the maximum spin-up of the signal over the whole tracking duration, $\dot{f}^{\rm max}_{\rm GW}$, we require
$\dot{f}^{\rm max}_{\rm GW} T_{\rm coh} \leq \Delta f$ following Eq.~\eqref{eqn:Tcoh_int}, and thus
\begin{equation}
    T_{\rm coh} \leq (2 \dot{f}^{\rm max}_{\rm GW} )^{-1/2}.
    \label{eqn:Tcoh}
\end{equation}
Since longer $T_{\rm coh}$ values yield better search sensitivity~\cite{Sun2018}, in a typical search we set $T_{\rm coh} = (2 \dot{f}^{\rm max}_{\rm GW} )^{-1/2}$ to maximize sensitivity. 
For vector boson signals, when the cloud is saturated at $t_{\rm sat}$, the strain amplitude reaches its peak, and the frequency evolution rate is also at its maximum, i.e., $\dot{f}^{\rm max}_{\rm GW} = \dot{f}_{\rm GW}(t_{\rm sat})$. 
Thus, we compute $\dot{f}^{\rm max}_{\rm GW}$ using \texttt{SuperRad} for each given system (e.g., Fig.~\ref{fig:fdot_v_M_v_chi}) and set $T_{\rm coh} = (2 \dot{f}^{\rm max}_{\rm GW} )^{-1/2}$, rounded down to the nearest 0.1 min. 

We estimate the likelihood of the signal in each frequency bin at each time step via the $\mathcal{F}$-statistic, which accounts for the Earth's motion with respect to the source through Doppler corrections~\cite{JKS, Prix2011}. We coherently integrate the data over the duration $T_{\rm coh}$; this results in \mbox{$N_T = T_{\rm obs}/T_{\rm coh}$} coherent $\mathcal{F}$-statistic segments over the total duration of the search $T_{\rm obs}$. The coherent segments are then combined incoherently using HMM tracking, as outlined in, e.g., Refs.~\cite{Suvorova2016, Sun2018}.
The particular choice of transition probability matrix, i.e., the probability for the signal frequency in each bin $f_i$ at the current time step to be in bin $f_j$ at the next time step, does not largely impact the sensitivity of the HMM tracking as long as it captures the general behavior
of the signal~\cite{Suvorova2016, Quinn2001}.
Given that the vector boson signal has a small positive $\dot{f}_{\rm GW}(t)$, i.e., the signal frequency slowly increases (see Sec.~\ref{sec:emission_&_timescales}), we apply for simplicity a uniform probability on $\dot{f}_{\rm GW}$ in the range of $[0, \dot{f}^{\rm max}_{\rm GW}]$ and write the transition probability $A_{f_j f_i}$ from one time step to the next as~\cite{Suvorova2016, Sun2018, Isi2019}
\begin{equation}
    A_{f_{i+1} f_i} = A_{f_i f_i} = \frac{1}{2},
    \label{eqn:transition_probability}
\end{equation}
with all other entries being zero. This means that, from the current time step to the next, a signal in bin $f_i$ either remains in bin $f_i$ or evolves to a higher frequency bin $f_{i+1}$.
A uniform prior $\Pi_{f_i} = N_Q^{-1}$ is applied on all frequency bins over the total frequency band being searched.
We use the Viterbi algorithm to solve the HMM and identify the most probable signal path recursively across the frequency-time plane~\cite{Viterbi1967}. 

In this work, we extend the standard HMM tracking used in CW searches to a much shorter timescale by introducing more flexible configurations and a new detection statistic. 
Reference~\cite{Isi2019} states that the $\mathcal{F}$-statistic/HMM-pipeline is not currently capable of tracking signals with $\dot{f}_{\rm GW} \gtrsim 10^{-8}$~Hz~s$^{-1}$. This statement makes the assumption that $T_{\rm SFT}$ used in the search is fixed to 30~mins (the standard SFT length used in CW searches). Due to the faster signal evolution characteristic of vector boson searches, both $T_{\rm coh}$ and $T_{\rm SFT}$ have to be much shorter [recall Eq.~\eqref{eqn:Tcoh}]. We set a lower cutoff $T_{\rm coh} \geq 1$~min with a minimum $T_{\rm SFT} \geq 0.25$~min. For each detector, the coherent $\mathcal{F}$-statistic calculation requires as input a minimum of two SFTs. However, to mitigate issues with insufficient data, we require at minimum a length of $4T_{\rm SFT}$ per $\mathcal{F}$-statistic segment. The cutoff on $T_{\rm coh}$ is chosen because we find that, for $T_{\rm coh} < 1$~min, the $\mathcal{F}$-statistic values computed in pure Gaussian noise no longer follow the expected central chi-squared distribution with four degrees of freedom. (For a more detailed discussion of the statistical studies behind short-segment $\mathcal{F}$-statistics, see Ref.~\cite{Covas2022}.) This cutoff corresponds to $\dot{f}_{\rm GW} = 1.39 \times 10^{-4}$~Hz~s$^{-1}$, which is the maximum $\dot{f}_{\rm GW}$ covered by the search method described in this paper. For signals that evolve more rapidly, other HMM-based methods outlined in Refs.~\cite{Sun2019, Banagiri2019} to track long-transient signals could be applied, but this is outside the scope of this study. 
On the other hand, for a parameter space that corresponds to much longer signals on timescales similar to typical CWs, we set an upper bound of $T_{\rm SFT}=30$~min to prevent power leakage due to the Doppler effect as the Earth rotates. We also set an upper bound of $T_{\rm coh}=10$~d for computing efficiency\footnote{The computing cost scales as $\sim T_{\rm coh}^2$ in HMM-based searches~\cite{Sun2018}.}, and to allow for some model uncertainty~\cite{Suvorova2016,Sun2018,Isi2019}.

In many of the existing HMM-based CW searches, a detection statistic called the ``Viterbi score'' is used to quantify the significance of the most probable path returned by the tracking (e.g., Refs.~\cite{Sun2018, Isi2019, O2_CygnusX1_scalar_bosons}). The Viterbi score $S$ is defined such that the log likelihood of the optimal Viterbi path equals the mean log likelihood of all $N_Q$ paths ending in $N_Q$ bins plus $S$ standard deviations at the final step $N_T$. In other words, the significance of the signal is evaluated by comparing the optimal path to all other paths in a given sub-band searched. We do not use the Viterbi score as our detection statistic in this study, however, because the Viterbi score is only reliable for $N_Q \gg N_T$. When $N_Q \sim N_T$, the $N_Q$ paths partially overlap and so are correlated~\cite{Millhouse2020}. Because the typical timescale of a vector boson signal is much shorter than the standard CW signals, most configurations in this study require wider frequency bins and fewer tracking steps compared to a standard CW search, and thus we often have $N_Q \gtrsim N_T$.

Instead, we define a new detection statistic as the total log-likelihood $\mathcal{L}$ of the optimal path divided by the number of steps $N_T$ written as
\begin{equation}
    \bar{\mathcal{L}} \equiv \mathcal{L}/N_T.
    \label{eqn:detection_statistic}
\end{equation}
In theory, we could simply use $\mathcal{L}$ as our detection statistic, as in Refs.~\cite{Millhouse2020, Beniwal2021, SNR_O3a, AMXP_O3, ScoX1_O3}. However, since we need to cover a wide parameter space for a given source (see Sec.~\ref{sec:param_space}), using $\bar{\mathcal{L}}$ allows us to remove the dominant dependence of the detection statistic on $N_T$ and generalize the detection statistic to many configurations covering a wide range of signal durations (note that $\bar{\mathcal{L}}$ still weakly depends on $N_T$ and $T_{\rm coh}$; see Sec.~\ref{sec:simulations}).
This is of particular importance for setting a detection threshold for a search. 

\subsection{Simulations and search configurations}
\label{sec:simulations}

\begin{figure}[hbt!]
	\centering
	\includegraphics[scale=.45]{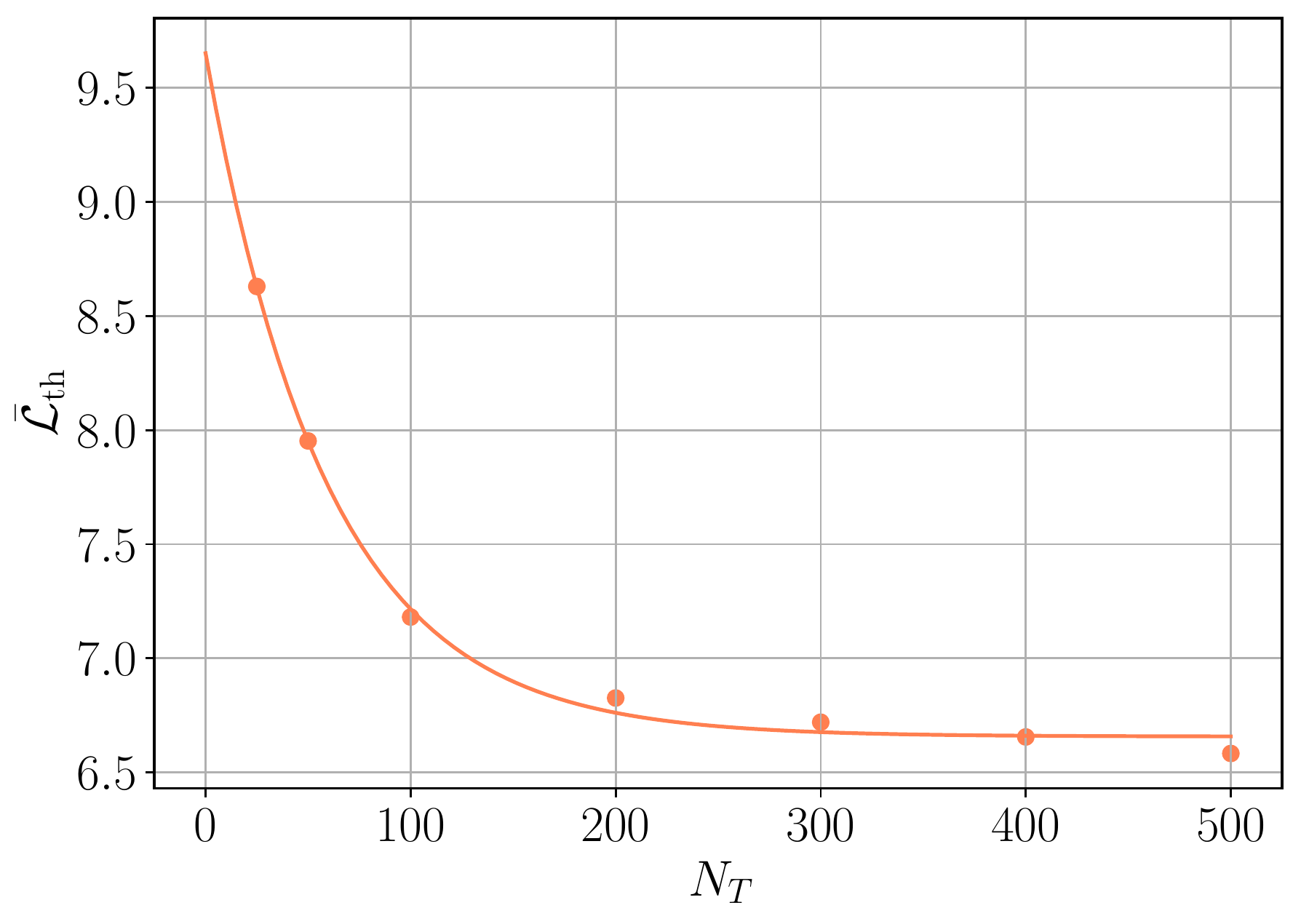}
	\caption{$\bar{\mathcal{L}}_{\rm th}$ as a function of $N_T$ for $T_{\rm coh} = 30$~min (data points are taken at $N_T$ = 25, 50, 100, 200, 300, 400, and 500~steps). The solid curve is an exponential decay fit: $\bar{\mathcal{L}}_{\rm th} = ae^{-b N_T} + c$ with fit parameters $a= 2.99$, $b = 0.0168$, and $c = 6.66$.}
	\label{fig:threshold_scaling}
\end{figure}

We define a 1\% false alarm probability threshold in each sub-band searched in order to quantify the confidence in a detected signal. Because we consider a wide range of search configurations in this study, each with a unique threshold depending on the choices of $T_{\rm coh}$ and $N_T$, we adopt a hybrid method to estimate thresholds based on both empirical simulations and analytical fitting. 
The procedure is as follows.
We first empirically test the following $T_{\rm coh}$ values: 1~min, 2~min, 5~min, 10~min, 30~min, 1~hr, 12~hr, 1~d, 5~d, and 10~d, covering the whole range of typical vector boson searches. 
For each choice of $T_{\rm coh}$, we consider five to eight $N_T$ values we might use in a real search. For each combination of $T_{\rm coh}$ and $N_T$, we obtain the threshold by running 300 searches in pure Gaussian noise within a single 1-Hz sub-band. We extract $\bar{\mathcal{L}}$ at the 99th percentile, denoted as $\bar{\mathcal{L}}_{\rm th}$. We consider anything with $\bar{\mathcal{L}} > \bar{\mathcal{L}}_{\rm th}$ to be a GW candidate (i.e., there is a 1\% probability the candidate is a false alarm in each sub-band). 
Then, for a given $T_{\rm coh}$, we plot $\bar{\mathcal{L}}_{\rm th}$ as a function of $N_T$ and fit an exponential decay curve to the data points; an example is shown in Fig.~\ref{fig:threshold_scaling} for $T_{\rm coh} = 30$~min for seven sample $N_T$ values. 
Repeating this process for all $T_{\rm coh}$ values chosen above, the fitted curves for different $T_{\rm coh}$ values end up roughly overlapping (the variation of $\bar{\mathcal{L}}_{\rm th}$ for any given $N_T$ is within $\sim \pm 5$\%), demonstrating that $\bar{\mathcal{L}}_{\rm th}$ depends more on $N_T$ than on $T_{\rm coh}$. 
It is indeed expected that the threshold is almost independent of $T_{\rm coh}$ because the $\mathcal{F}$-statistic values computed over each coherent step in pure Gaussian noise should follow the same central chi-squared distribution with four degrees of freedom.
Due to the maximization in the HMM tracking, the tail in the $\bar{\mathcal{L}} = \mathcal{L}/N_T$ distribution still depends on $N_T$. 
In practice, we determine a choice of ($T_{\rm coh}$, $N_T$) for the search based on the signal parameter space. We then find $\bar{\mathcal{L}}_{\rm th}$ at the chosen $N_T$ from the exponential fitting curve.
Although the deviations among the fitting curves for different $T_{\rm coh}$ values are small, we take the curve obtained with the $T_{\rm coh}$ value (among the 10 values tested) closest to the one chosen for the search configuration.
We use this method in this study in order to get the threshold and obtain an estimate of the search sensitivity across the whole parameter space while saving on computing costs (see Sec.~\ref{sec:horizon_distance}). In a real directed search, we can always empirically obtain the threshold using the specific search configurations suitable for a given source to avoid small statistical deviations introduced by interpolation.


Once we have obtained detection thresholds, we run searches for synthetic vector boson signals simulated based on the signal morphology described in Secs.~\ref{sec:emission_&_timescales}--\ref{sec:GW_signals} across the parameter space and demonstrate how well the method is able to recover the signal. 
As an example, we consider a system with $M_i = 200~M_{\odot}$, $\chi_i = 0.6$, and $\alpha = \alpha_{\rm opt} = 0.141$. We place this system at $d = 500$~Mpc, which corresponds to a peak strain amplitude of $\hpeak = 5.66 \times 10^{-25}$. We assume the system has the optimal orientation, i.e., $\cos{\iota} = 1.0$. We use \texttt{SuperRad} to build a signal waveform based on these parameters and inject the signal into Gaussian noise with an amplitude spectral density (ASD) of $S_h^{1/2} = 4 \times 10^{-24}$~Hz$^{-1/2}$ (the aLIGO design sensitivity in the most sensitive frequency band $\sim 10^2$~Hz) using the simulateCW Python module in the LALPulsar library of LALSuite~\cite{lalsuite, swiglal}. We inject the signal at RA = 4.41955~rad and Dec = 0.62385~rad in the 1-Hz band starting from 40~Hz (with two aLIGO detectors).

\begin{figure}[hbt!]
	\centering
	\includegraphics[scale=.44]{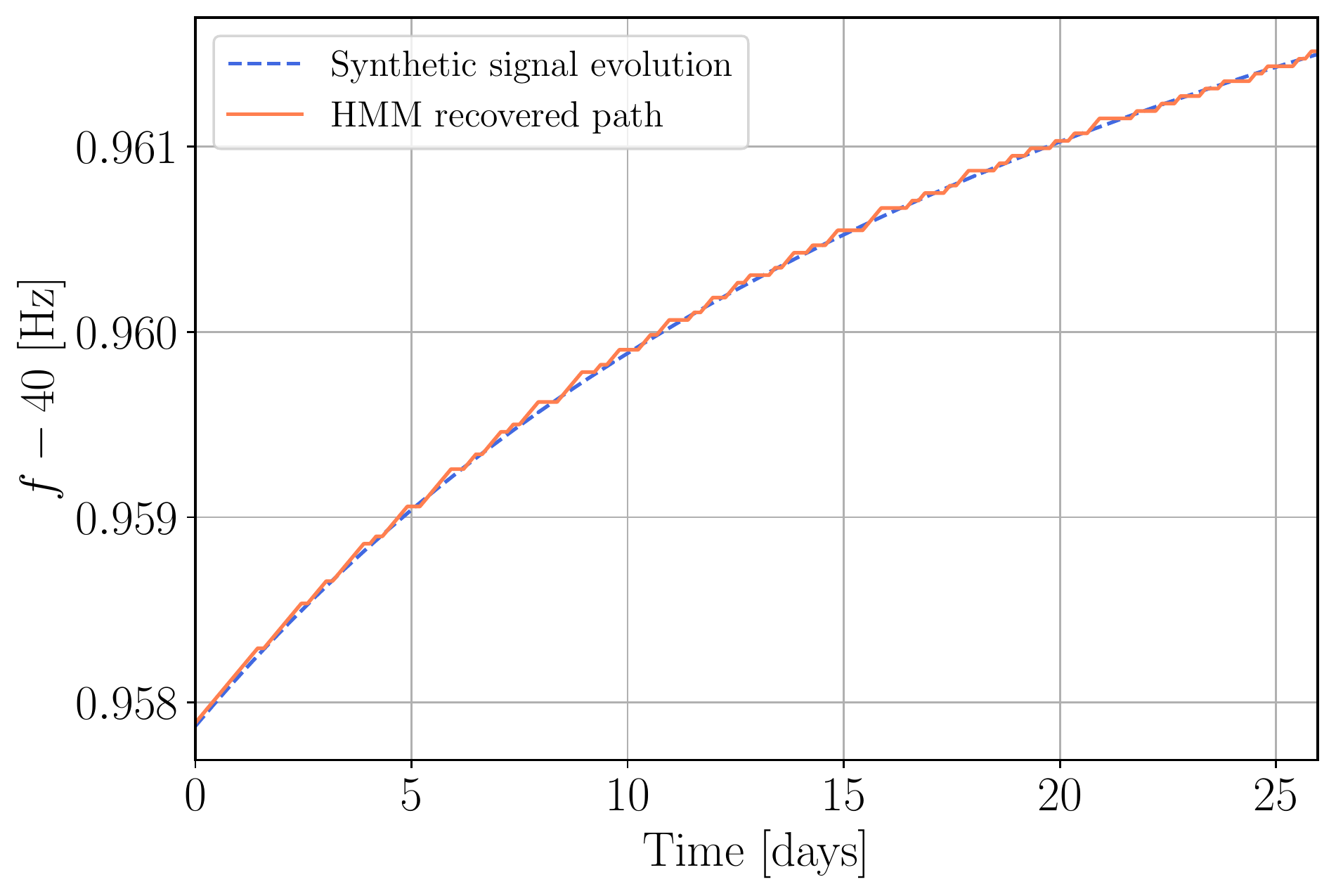}
	\caption{Viterbi tracking (solid orange curve) for a synthetic vector boson signal (dashed blue curve) injected into Gaussian noise with $S_h^{1/2} = 4 \times 10^{-24}$~Hz$^{-1/2}$ for two aLIGO detectors (system parameters: $M_i = 200~M_{\odot}$, $\chi_i = 0.6$, $\alpha_{\rm opt} = 0.141$, and $d= 500$~Mpc). We use $T_{\rm coh} = 207$~min = 3.45~hr and run the search for a total duration of $\sim 26$ d ($N_T = 181$~steps). The detection statistic is $\bar{\mathcal{L}} = 57.43 > \bar{\mathcal{L}}_{\rm th} = 6.73$.}
	\label{fig:Viterbi_tracking_1}
\end{figure}

Given the $\dot{f}_{\rm GW}$ estimated by \texttt{SuperRad}, we choose the best possible coherent length based on Eq.~\eqref{eqn:Tcoh}, $T_{\rm coh} = 207$~min, and track the signal over $\sim 26$~days. The result is shown in Fig.~\ref{fig:Viterbi_tracking_1}, with the injection indicated by the dashed blue curve and the signal path recovered by HMM indicated by the solid orange curve. The stairstep pattern of the recovered signal is a result of the search being divided into discrete frequency bins and time steps. The detection statistic associated with this recovered path is $\bar{\mathcal{L}} = 57.43$, well above the estimated threshold, $\bar{\mathcal{L}}_{\rm th} = 6.73$, indicating a successful detection. As demonstrated, the HMM is able to accurately reconstruct the signal down to a root-mean-square error of $2.12 \times 10^{-5}$~Hz.

As mentioned, we run the search over a total duration $T_{\rm obs} \approx 26$~days in the above example. Unlike standard CW signals, which have essentially constant strain amplitude over the entire observing time of $\sim$ years, vector boson signals decay much quicker. Thus, there is an optimal range for $T_{\rm obs}$ which is long enough to accumulate a significant SNR, but short enough to not accumulate pure noise after the signal strength falls below the detection limit.
The optimal range of $T_{\rm obs}$ varies for different systems but is expected to be on the order of $ \tau_{\rm GW}$ [see Eq.~\eqref{eqn:emission_timescale} and Fig.~\ref{fig:tauGW_v_M_v_chi}].
Hence, we use $T_{\rm obs} \approx \tau_{\rm GW}$ (with some rounding involved such that $T_{\rm obs}$ is evenly divided into $T_{\rm coh}$ intervals). 

\begin{figure*}[hbt!]
	\centering
	\includegraphics[scale=.6]{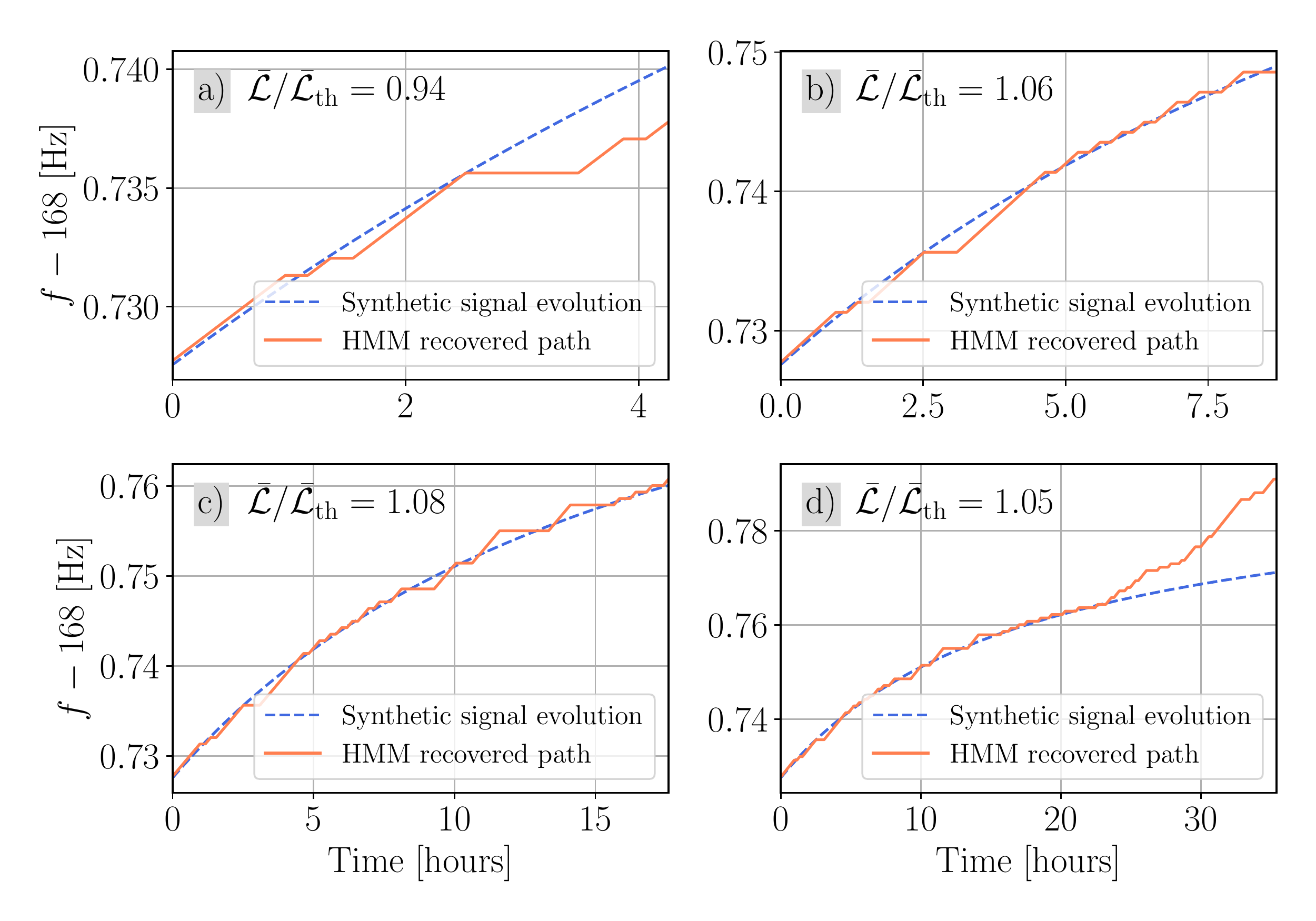}
	\caption{Viterbi tracking (solid orange curve) for a synthetic vector boson signal (dashed blue curve) injected into Gaussian noise with $S_h^{1/2} = 4 \times 10^{-24}$~Hz$^{-1/2}$ for two aLIGO detectors (system parameters: $M_i = 60~M_{\odot}$, $\chi_i = 0.7$, $\alpha_{\rm opt} = 0.176$, and $d = 500$~Mpc). We use $T_{\rm coh} = 11.6$~min and run the search for a) 23 steps, b) 46 steps, c) 92 steps, and d) 184 steps, corresponding to $T_{\rm obs} = 0.25 \tau_{\rm GW}$, $0.5 \tau_{\rm GW}$, $\tau_{\rm GW}$, and $2 \tau_{\rm GW}$, respectively ($\tau_{\rm GW} = 0.74$~days). In panels b), c), and d), $\bar{\mathcal{L}}$ is above the corresponding threshold, whereas in panel a), $\bar{\mathcal{L}} < \bar{\mathcal{L}}_{\rm th}$, which is a non-detection.}
	\label{fig:Viterbi_tracking_2}
\end{figure*}

To demonstrate the effect of searching over longer or shorter durations, we show another example in Fig.~\ref{fig:Viterbi_tracking_2}. We consider a system with $M_i = 60~M_{\odot}$, $\chi_i = 0.7$, and $\alpha = \alpha_{\rm opt} = 0.176$ at $d = 500$~Mpc ($\cos{\iota} = 1.0$) and inject a synthetic signal into Gaussian noise ($S_h^{1/2} = 4 \times 10^{-24}$~Hz$^{-1/2}$) in the 168--169~Hz sub-band for two aLIGO detectors. 
We track the injection with $T_{\rm coh} = 11.6$~min for $N_T = 23$, 46, 92, and 184~steps; the respective trackings are shown in panels a)--d), corresponding to $T_{\rm obs} = 0.25 \tau_{\rm GW}$, $0.5 \tau_{\rm GW}$, $\tau_{\rm GW}$, and $2 \tau_{\rm GW}$, respectively. 
In panel a), we find that $\mathcal{\bar{L}}$ falls below the threshold, so the signal is not recovered. This is because $T_{\rm obs}$ is too short to accumulate enough signal power. In panels b) and c), we have $\mathcal{\bar{L}} > \mathcal{\bar{L}}_{\rm th}$, so the signal is successfully recovered in both cases. The recovered signals in each panel (solid orange curves) align well with the injected signals (dashed blue curves). As such, $T_{\rm obs} = 0.5 \tau_{\rm GW}$ and $\tau_{\rm GW}$ are both good choices for this system. In panel d), while $\mathcal{\bar{L}}$ is above the threshold, it is only marginally so. This is because $h_0$ decreases as the boson cloud dissipates, and as shown in d), the tracking loses the signal and begins to collect pure noise in the last third of the total $T_{\rm obs}$, resulting in a less significant detection statistic.
Overall, we find it is safest to use $T_{\rm obs} \approx \tau_{\rm GW}$, which always falls in the optimal range for the systems we have tested across the system parameter space.

Here we have considered the optimal $T_{\rm obs}$ range for a marginal signal in order to quantify the search sensitivity. If the signal is sufficiently loud, using $T_{\rm obs} \approx \tau_{\rm GW}$ is still safe for detecting the signal, but extending $T_{\rm obs}$ would further increase the SNR, allowing a follow-up verification for the signal candidate.

\section{Search sensitivity and horizon distance}
\label{sec:sensitivity}

Based on the simulations described above, we estimate horizon distances in optimal scenarios for current and future generation detectors in Sec.~\ref{sec:horizon_distance} and discuss the non-optimal cases in Sec.~\ref{sec:non-optimal}. 
Because we do not make any assumptions about the origins of our target sources, our conclusions are broadly applicable to stellar-mass black holes with reasonably well-constrained sky positions and intrinsic parameters.

\subsection{Horizon distance estimate}
\label{sec:horizon_distance}

In this section, we quantify the horizon distance $d_H$, defined as the
farthest luminosity distance we would be able to detect a vector boson signal from a
given black hole in the optimal scenario. Here, the optimal scenario is defined
as: i) the boson mass optimally matches its host black hole in terms of
maximizing the intrinsic strain amplitude when the cloud is saturated, and ii)
the black hole-boson system is optimally oriented (face-on or face-off), such
that the effective strain amplitude on Earth is maximized. 

In Sec.~IV~B of Ref.~\cite{Isi2019}, the authors estimate horizon distances for scalar clouds by first obtaining the search sensitivity on signal strain amplitude corresponding to 95\% detection efficiency at 1\% false-alarm probability, denoted by $h_0^{95\%}$, for a particular search configuration. Sensitivities under other search configurations (i.e., different choices of $T_{\rm coh}$ and $T_{\rm obs}$) can be obtained by the following scaling~\cite{Sun2018}
\begin{equation}
    h_0^{95\%}(f) \propto \frac{S_h(f)^{1/2}}{N_{\rm ifo}^{1/2} \left( T_{\rm coh} T_{\rm obs} \right)^{1/4}},
    \label{eqn:sensitivity_scaling}
\end{equation}
assuming that $N_{\rm ifo}$ detectors in the network have the same ASD at the signal frequency. 
The horizon distance is then the luminosity distance of the system at which the signal strain $h_0$ at $d_H$ equals $h_0^{95\%}(f_{\rm det})$, where $f_{\rm det}$ is the signal frequency in the detector frame. 

We do not follow this scaling in this study, however, because Eq.~\eqref{eqn:sensitivity_scaling} is not as reliable for short signals. Moreover, the effects of redshift in vector boson searches are more significant since we can reach much farther into the Universe, as we discuss below. 
Thus, we need to consider a wide range of possible values of $h_0$, $f_{\rm GW}$, $T_{\rm coh}$, $T_{\rm SFT}$, and $T_{\rm obs}$ for a given system, all depending on the system's luminosity distance from Earth. Because of the challenges these factors pose, we instead estimate the horizon distance directly on a grid of black hole masses and spins.

\begin{figure}[hbt!]
	\centering
	\includegraphics[scale=.45]{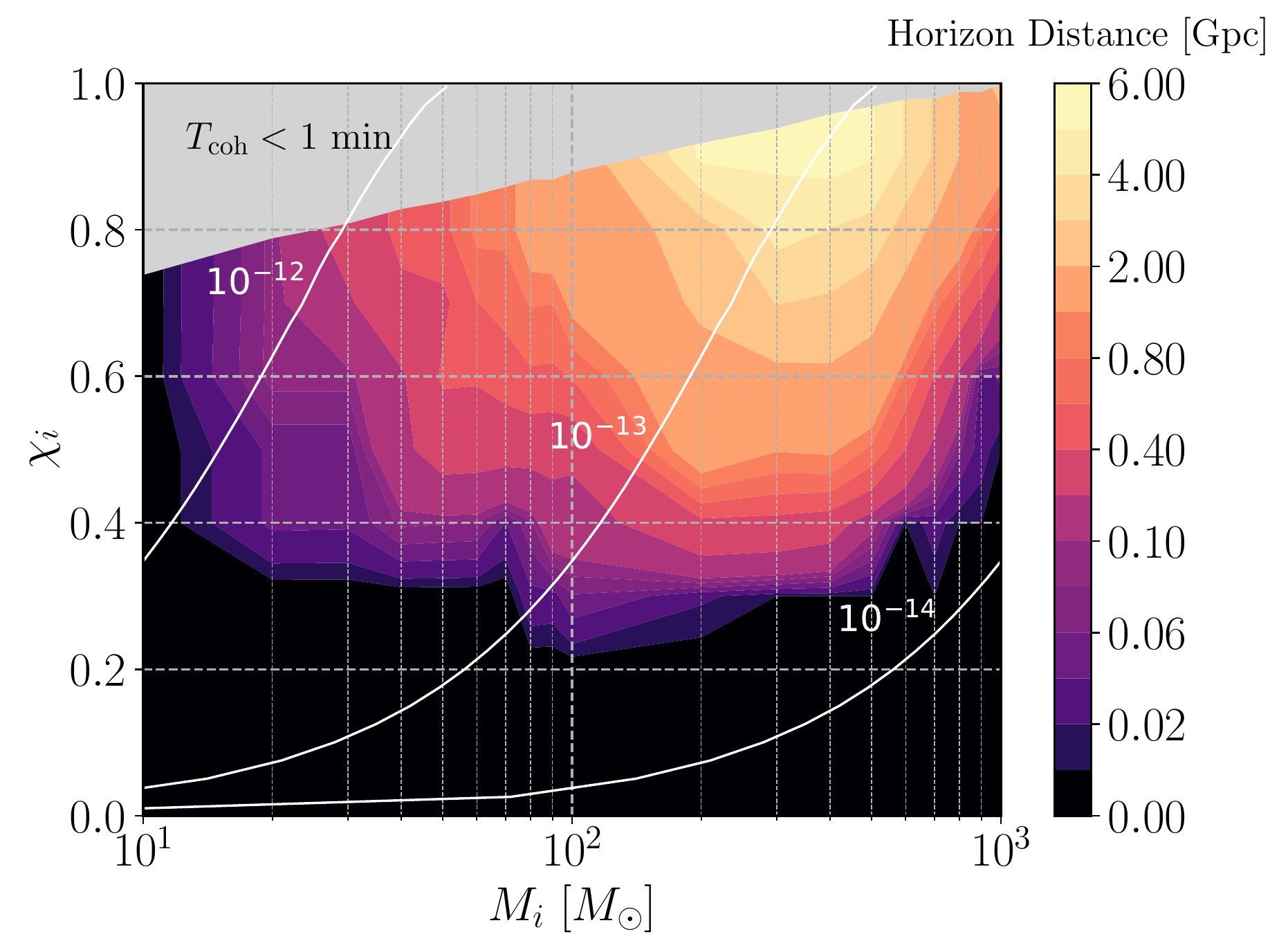}
	\caption{Horizon distance (colored contour) as a function of the initial black hole mass $M_i$ and initial spin $\chi_i$ for two aLIGO detectors at design sensitivity. The gray region marks the parameter space where the signal is evolving too quickly to be tracked using the method in this paper ($\dot{f}_{\rm det} > 1.39 \times 10^{-4}$ Hz\,s$^{-1}$). The white contours mark the optimally matched boson masses [as defined in Eq.~\eqref{eq:alphaopt}] in eV.}
	\label{fig:horizon_distances}
\end{figure}

Figure~\ref{fig:horizon_distances} shows the estimated horizon distances as a function of $M_i$ and $\chi_i$, assuming a network of two aLIGO detectors at design sensitivity~\cite{asd2020}. Here, the results are presented for the optimal scenario described above (i.e., the boson mass is $m^{\rm opt}_V=\alpha_{\rm opt}\hbar/r_g$ and the system is face-on/face-off).
We limit $T_{\rm coh}$ to $1~{\rm min} \leq T_{\rm coh} \leq 10$~days. (See Sec.~\ref{sec:search_methods} for the justification.) 
We set the total observing time \mbox{$T_{\rm obs} = \min (\tau_{\rm GW}, 180~{\rm d})$}. (See Fig.~\ref{fig:tauGW_v_M_v_chi} for the typical range of $\tau_{\rm GW}$ values for a given black hole.) We select a 180~d cutoff for the total observing time because we aim to follow up promising CBC merger remnants in LIGO-Virgo-KAGRA (LVK) observing runs, which usually last $\sim 1$ year with events detected throughout the run. This 180~d cutoff is also motivated by the need to save on computing costs wherever possible.
The gray area in the top left corner of Fig.~\ref{fig:horizon_distances} denotes the region of the parameter space where the maximum allowed $T_{\rm coh}$ is shorter than 1~min and the signal is evolving too quickly for this method to cover.\footnote{As discussed in Sec.~\ref{sec:search_methods}, alternative HMM-based methods, e.g., Refs.~\cite{Sun2019, Banagiri2019}, can be used for rapidly evolving signals in the gray region. Also see Sec.~\ref{sec:non-optimal} for additional discussion regarding non-optimally matching scenarios.} Moreover, because aLIGO detectors have little sensitivity below $\sim 5$~Hz, we set a lower cutoff in frequency at 5~Hz. This results in a noticeable suppression in the horizon distance at very large $M_i$, where the systems are optimal for low-mass bosons and tend to emit at lower frequencies.

The effect of redshift $z$ on the signal frequency is non-negligible at large luminosity distances and can be expressed as $f_{\rm det} = f_{\rm src} (1+z)^{-1}$ and $\dot{f}_{\rm det} = \dot{f}_{\rm src} (1+z)^{-2}$, where $f_{\rm src}$ ($\dot{f}_{\rm src}$) is the frequency (derivative) in the source frame, and $f_{\rm det}$ ($\dot{f}_{\rm det}$) is the respective quantity in the detector frame. This allows us to use longer coherent lengths, $T_{\rm coh}^z = T_{\rm coh}(1+z)$, and longer SFT lengths, $T_{\rm SFT}^z = T_{\rm SFT}(1+z)$, where the superscript $z$ denotes redshifted signals. 
Similarly, we have a redshifted GW emission timescale $\tau_{\rm GW}^z = \tau_{\rm GW}(1+z)$ and thus are able to observe over longer durations $T_{\rm obs}^z = T_{\rm obs} (1+z)$. Hence, for a given system at cosmological distances, the search sensitivity may improve as distance increases because we are able to extend $T_{\rm coh}$ (and search sensitivity improves as $T_{\rm coh}$ increases); the search sensitivity may also degrade, however, since the distance to the system is increasing (and the signal amplitude linearly scales with the inverse of the distance). Whether it is a net gain or loss in sensitivity depends on the configuration of the system, redshift, and the detector noise ASD at the signal frequency in the detector frame. 

The estimated horizon distances shown in Fig.~\ref{fig:horizon_distances} have the redshift effects taken into account. The procedure to account for redshift is as follows. 
For a given black hole-boson system (a given set of $M_i$, $\chi_i$, and $m_V^{\rm opt}$), we first inject synthetic signals calculated by \texttt{SuperRad} into Gaussian noise with the following set of extrinsic parameters: $d = 50$~Mpc, $\cos{\iota} = 1$, a randomized polarization angle, a fixed ASD of $S_h^{1/2} = 4 \times 10^{-24}$~Hz$^{-1/2}$, and a set of arbitrarily chosen sky coordinates ${\rm (RA, Dec)} = (4.41955, 0.62385)$~rad. 
We choose the optimal search configuration for the system that is assumed to lie at this distance and attempt to recover the signal using HMM. If the signal is recovered with $\bar{\mathcal{L}} > \bar{\mathcal{L}}_{\rm th}$, we increase $d$ to 100~Mpc and repeat the same process. We continue to increase the distance by an interval of 100~Mpc until $\bar{\mathcal{L}}$ drops below the threshold. We quote the largest distance at which we are still able to recover the signal as the horizon distance, $d_H$, for each given black hole in the $(M_i, \chi_i)$ plane. Then we rescale $d_H$ based on the frequency-dependent ASD curve for aLIGO design sensitivity~\cite{asd2020}, with $T_{\rm coh}$, $T_{\rm obs}$, and the redshift effect all taken into account, following the scaling given by Eq.~\eqref{eqn:sensitivity_scaling}:
\begin{eqnarray}
    \nonumber
    &&\frac{h_0(d_H) \left[ T_{\rm coh}(d_H) T_{\rm obs}(d_H) \right]^{1/4}}{S_h^{1/2}} \\
    &=& \frac{h_0(d'_H) \left[ T_{\rm coh}(d'_H) T_{\rm obs}(d'_H) \right]^{1/4}}{S_h^{1/2}(d'_H)},
    \label{eqn:scale_dH}
\end{eqnarray}
where $S_h^{1/2}$ is fixed to the value used in the simulations, i.e., $4 \times 10^{-24}$~Hz$^{-1/2}$, $S_h^{1/2}(d'_H)$ is the aLIGO design ASD at the redshifted signal frequency $f_{\rm det}$ (which in turn depends on $d_H$), and $d'_H$ is the target horizon distance scaled to the aLIGO design sensitivity. We obtain the target $d'_H$ value by numerically solving Eq.~\eqref{eqn:scale_dH} for each system.

As expected, the horizon distance generally increases with both $M_i$ and $\chi_i$. High-mass black holes lead to signals with smaller $\dot{f}_{\rm GW}$ values, allowing us to use longer $T_{\rm coh}$ segments, yielding increased sensitivity. 
The gain is diminished by the fact that the lower-mass bosons matching the higher-mass black holes emit at lower frequencies, where ground-based detectors are less sensitive due to seismic noise. When the horizon distances correspond to high redshifts, the signals are redshifted to even lower frequencies. Hence, the horizon distance degrades towards the higher end of the black hole mass spectrum in the figure.
Towards the lower end of the $M_i$ spectrum, the optimally matching bosons have higher mass and emit at higher frequencies, where the detector's sensitivity is limited by shot noise. In addition, boson clouds around smaller black holes emit lower-amplitude GWs.
Thus in the low $M_i$ region, the search sensitivity is also limited. 
Nevertheless, unlike the expected signals generated by scalar clouds around CBC remnants, for which the detection prospects are dim for current generation detectors,\footnote{Scalar clouds emit weaker signals that occur over much longer timescales; in this case, galactic black holes that are older but more nearby would be the more promising targets for current generation detectors (see, e.g., Refs.~\cite{O3_all-sky_scalar_bosons, O2_CygnusX1_scalar_bosons}). However, it is important to note that when targeting unknown black holes and/or known black holes with unknown ages within our galaxy, constraints derived on the boson mass are contingent on the assumed system age as well as the black hole population.} for a parameter space with $M_i \gtrsim 60 M_\odot$ and $\chi_i \gtrsim 0.6$ (corresponding to a boson mass of $\sim 10^{-13}$~eV and $d \gtrsim 400$~Mpc), searches for vector boson signals are promising using existing detectors; CBC events detected in previous observing runs are found at luminosity distances $\lesssim \mathcal{O}$(1~Gpc).

\begin{figure}[hbt!]
	\centering
	\includegraphics[scale=.53]{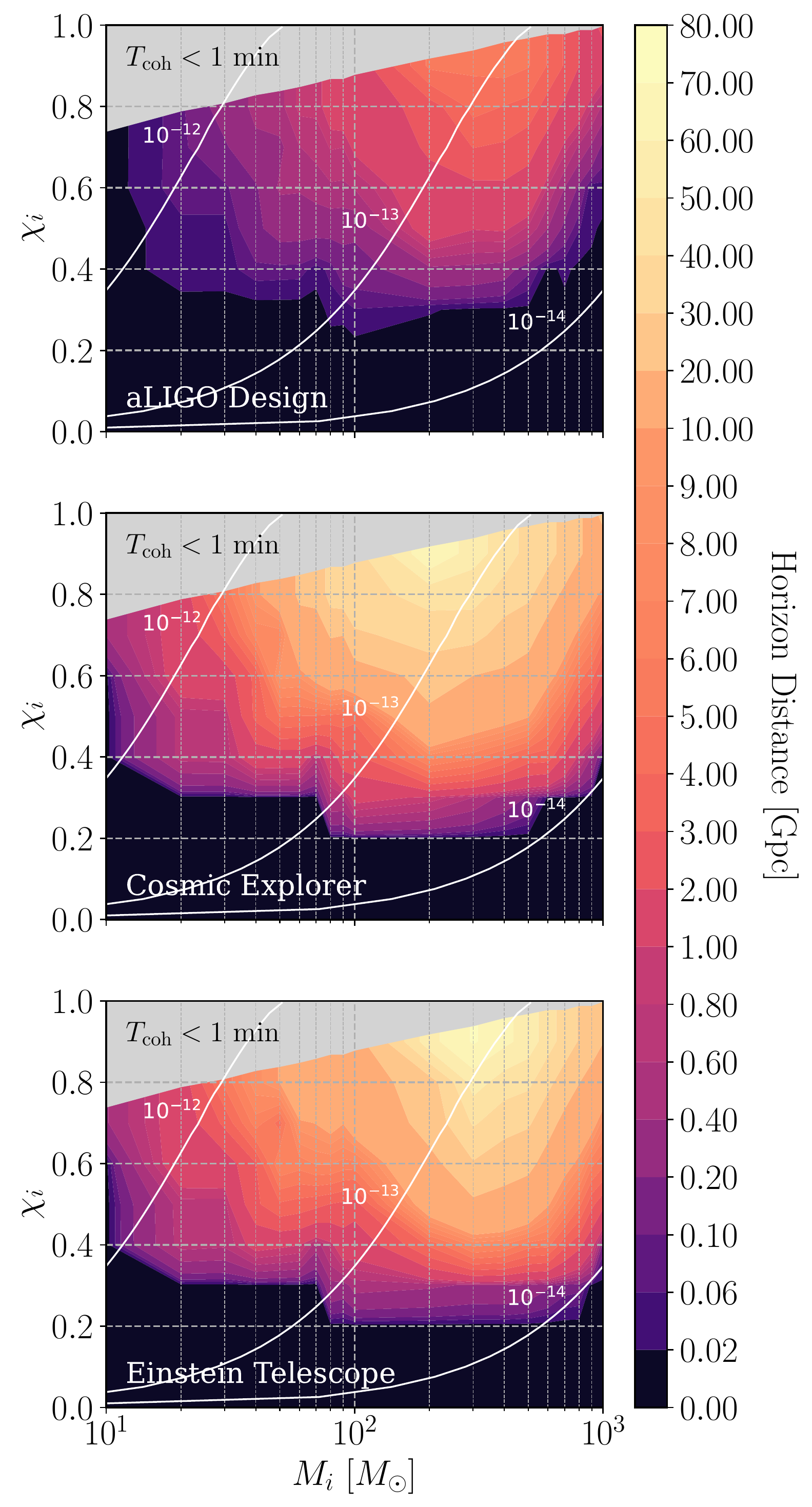}
	\caption{Horizon distance (colored contour) as a function of the initial black hole mass $M_i$ and initial spin $\chi_i$ for two aLIGO detectors at design sensitivity (top), one Cosmic Explorer (middle), and one Einstein Telescope with three identical observatories at the same triangular site (bottom). The gray region marks the parameter space where the signal is evolving too quickly to be tracked using the method in this paper ($\dot{f}_{\rm det} > 1.39 \times 10^{-4}$ Hz\,s$^{-1}$). The white contours mark the optimally matched [as defined in Eq.~\eqref{eq:alphaopt}] boson masses (in eV), roughly indicating the parameter space that can be probed with these ground-based detectors. 
    }
	\label{fig:horizon_distances_3x}
\end{figure}

In Fig.~\ref{fig:horizon_distances_3x}, we compare the horizon distances using two aLIGO detectors at design sensitivity with the proposed next-generation detectors: Cosmic Explorer~\cite{Abbott-CE, Evans:2021gyd, Reitze2019cosmic} and Einstein Telescope~\cite{Punturo:2010zz, Hild:2010id, et-design, Maggiore2020}. The top panel is the same as Fig.~\ref{fig:horizon_distances}, but with a different color scale for visual comparison with the bottom two panels. 
We use the same method as described above to rescale the horizon distances using the design ASD curves for Cosmic Explorer and Einstein Telescope (also with a lower cutoff frequency at 5~Hz)~\cite{asd2020, CEasd2022}.
According to the figure, future generation detectors will improve the horizon distances by about an order of magnitude, allowing us to probe a much wider parameter space for boson masses $\sim 10^{-14}$--$10^{-12}$~eV. 

For comparison, we calculate the matched filter SNR (SNR$_{\rm mf}$) for each of three example black holes with optimally matched boson masses at the aLIGO horizon distances (Table~\ref{tab:matched_filter_SNR}). The SNR$_{\rm mf}$ values are in the range $\approx 15$--25, roughly what we would expect for detection in a semicoherent HMM search. In Ref.~\cite{Chan2022}, it is assumed that any signal with SNR$_{\rm mf} > 8$ can be detected, and correspondingly, they find horizon distances that are a factor of a few larger than found here. (Reference~\cite{Chan2022} also uses a non-relativistic estimate of the GW amplitude.) In reality, a more flexible method that is less susceptible to model uncertainties, like the one described in this paper, sacrifices some sensitivity and requires a higher SNR for confident detection. That is, for a less sensitive semicoherent search, we require a signal with higher SNR (SNR$_{\rm mf} \gtrsim 15$--25) than that which is required in a fully coherent search (SNR$_{\rm mf} > 8$) to ensure the signal is detectable.

%
\begin{table}[tbh]
	\centering
	\setlength{\tabcolsep}{15pt}
	\renewcommand\arraystretch{1.2}
	\begin{tabular}{lccc}
		\hline
		$M_i~[M_{\odot}]$ & $\chi_i$ & $d_H$~[Gpc] & ${\rm SNR}_{\rm mf}$ \\
		\hline
        40 & 0.5 & 0.175 & 25.3 \\
        80 & 0.6 & 0.564 & 20.7 \\
		100 & 0.7 & 1.096 & 14.9 \\        
		\hline
	\end{tabular}
	\caption{Matched filter SNRs for three sample systems at the estimated horizon distances using two aLIGO detectors.}
	\label{tab:matched_filter_SNR}
\end{table}

\subsection{Non-optimal scenarios}
\label{sec:non-optimal}

\begin{figure}[hbt!]
	\centering
	\includegraphics[scale=.45]{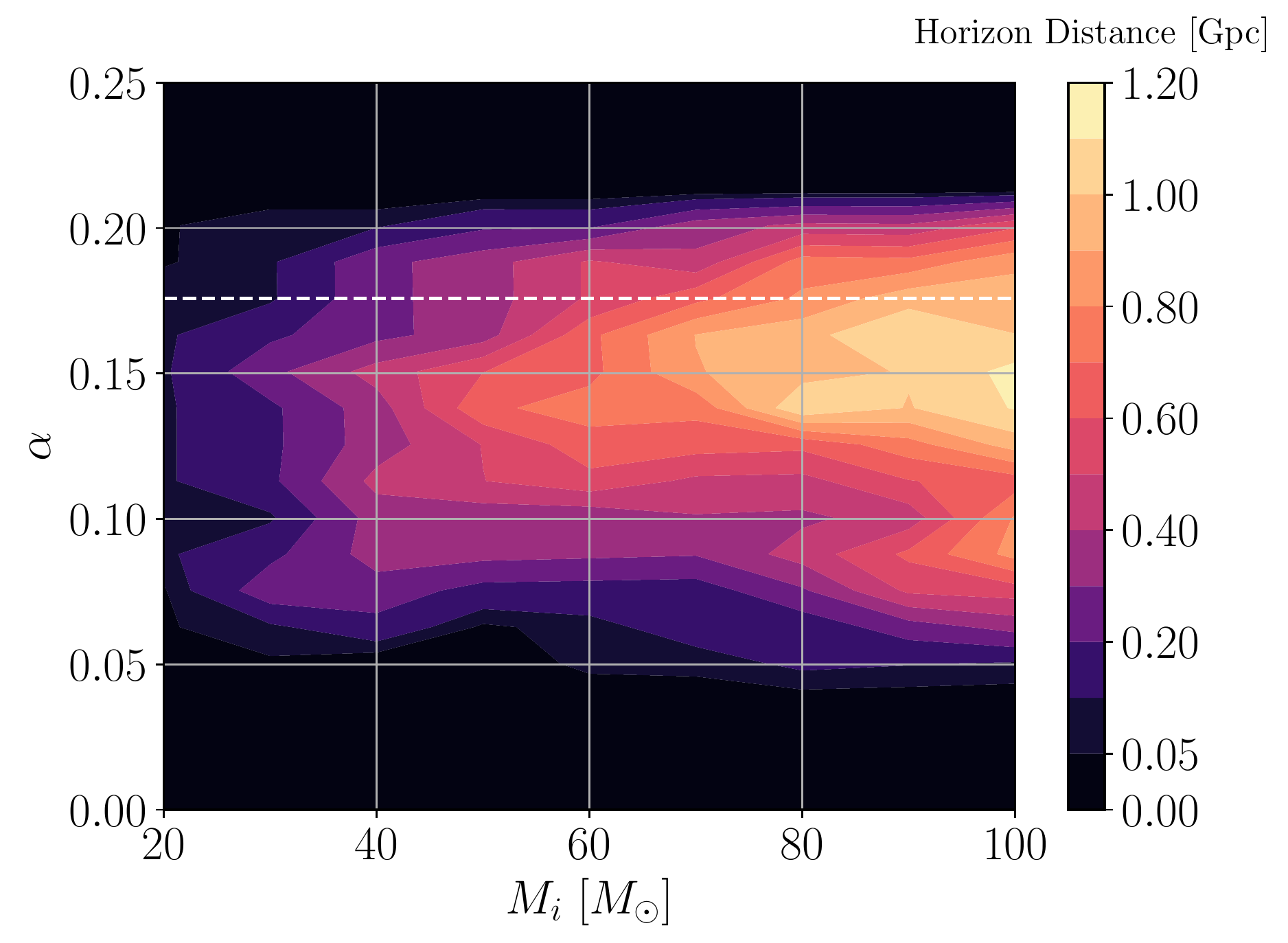}
	\caption{Horizon distance (colored contour) as a function of $M_i$ and $\alpha$ (at $\chi_i = 0.7$) for two aLIGO detectors at design sensitivity. The dashed white line marks $\alpha_{\rm opt} = 0.176$, the $\alpha$-value corresponding to the optimally matched boson mass for each black hole mass.}
	\label{fig:horizon_distances_non-optimal}
\end{figure}

Up until this point, all horizon distances have been estimated using the optimally-matched boson mass $m_V^{\rm opt}$. 
In the case of scalar bosons, the optimally matching case automatically yields the maximum horizon distance~\cite{Isi2019}, assuming no impact from the detector ASD, since the signals last for timescales on the order of years or more and the signal strain is maximized over the whole observing time. 
This is not always the case for vector bosons, as we demonstrate in Fig.~\ref{fig:horizon_distances_non-optimal}, which shows the horizon distances for a range of $\alpha$ values as a function of $M_i$ for a fixed $\chi_i = 0.7$. 
The horizontal dashed line marks $\alpha_{\rm opt}=0.176$. We see that $\alpha_{\rm opt}$ does not align with the maximum horizon distance for any given $M_i$; rather, the maximum $d_H$ (with a factor of $\sim 1.2$--2 improvement) lies roughly at $\alpha \approx 0.15$ with a long tail into the lower $\alpha$ values.
This behavior is unique to vector boson signals, which are much shorter than scalar signals.

The optimally matching value of $\alpha$, by construction, has the maximum strain when the cloud is saturated.
However, since this means the radiated power will be nearly maximized, the signal will evolve rapidly with a large $\dot{f}_{\rm GW}$ and a short $\tau_{\rm GW}$, which limits the length of $T_{\rm coh}$ and $T_{\rm obs}$ that can be used in the search and thus degrades the sensitivity.
On the other hand, when we consider a suboptimal boson mass for a given black hole, by Eqs.~\eqref{eqn:GW_power_approx} and~\eqref{eqn:emission_timescale_approx}, the cloud radiates at lower power and emits GWs over a longer timescale. Although the signal strain is smaller due to lower intrinsic GW power, because the signal evolves more slowly and lasts longer, we are able to extend $T_{\rm coh}$, gaining sensitivity, and we can track over a longer $T_{\rm obs}$, accumulating a higher SNR. In Fig.~\ref{fig:horizon_distances_non-optimal}, when $\alpha \lesssim \alpha_{\rm opt}$, the gain in sensitivity outweighs the loss due to a smaller signal strain. But as $\alpha$ further decreases, the signal becomes too weak, and the sensitivity degrades again.
Hence, the horizon distances presented in Sec.~\ref{sec:horizon_distance} are only for the optimally matching boson for each black hole; they are not necessarily the largest luminosity distances we can reach for any possible boson mass. Some suboptimal boson masses will lead to better detection prospects.
It follows that for a given black hole, it is not only possible, but also beneficial for us to probe a range of boson masses.


Although not shown in Fig.~\ref{fig:horizon_distances_non-optimal}, as $\chi_i$ increases, the $\alpha$ value corresponding to the maximum $d_H$ shifts more significantly from $\alpha_{\rm opt}$. This is because for higher-spin black holes with $\alpha = \alpha_{\rm opt}$, the signal frequency evolves quicker and thus requires shorter $T_{\rm coh}$ lengths in the search; the sensitivity gain at non-optimal $\alpha$ values, which allow for longer $T_{\rm coh}$ segments, is then more significant.

The gray shaded regions in Figs.~\ref{fig:horizon_distances} and~\ref{fig:horizon_distances_3x}, which mark the parameter space where the signals evolve too quickly to track with the method described in this paper, are not necessarily inaccessible. For suboptimal $\alpha$ values, the frequency derivative is smaller, allowing us to significantly extend $T_{\rm coh}$ and probe the gray region of the parameter space for such boson masses.

We also consider the case in which the source is not optimally oriented with respect to the detectors, i.e., $\cos{\iota} \neq \pm 1$. The luminosity distance ($d$) and orientation ($\iota$) of the source are degenerate, and we can write the effective strain amplitude seen by the detectors as~\cite{JKS, Jones2022}
\begin{equation}
    h_0^{\rm eff} = h_0(d)~2^{-1/2} \{[(1+\cos^2{\iota})/2]^2 + \cos^2{\iota}\}^{1/2}.
    \label{eqn:h0_eff}
\end{equation}
Since the sensitivity to the effective strain $h_0^{\rm eff}$ remains fixed within a given detector, we can analytically scale the horizon luminosity distance for a non-optimally oriented system using Eq.~\eqref{eqn:h0_eff}.\footnote{The scaling in Eq.~\eqref{eqn:h0_eff} is an approximation and only becomes exact in the non-relativistic limit ($\alpha \ll 1)$. However, it is a good approximation for all systems considered in this study. (See the discussion below Eq.~\eqref{eqn:h_cross} for further details.) In addition, Eq.~\eqref{eqn:h0_eff} assumes a randomized polarization angle and neglects the weak impact from the sky position in this scaling.}

\section{Sources and sky localization}
\label{sec:sources}
As discussed in Sec.~\ref{sec:intro}, although constraints have already been placed on the boson mass using black hole spin measurements, there are significant associated uncertainties. Searches targeting individual black holes represent a more direct approach to testing the superradiance phenomenon and constraining the boson mass.
We describe promising search targets for vector bosons in Sec.~\ref{sec:targets}. Then, we discuss the impact of the sky localization of the target black hole and analyze two different systems as examples in Sec.~\ref{sec:sky_grid}.

\subsection{CBC remnant black holes}
\label{sec:targets}

To determine what types of black holes would be ideal targets for vector boson searches,
we first consider black holes with well-estimated masses, spins, and ages that could host boson clouds whose signals would fall within reach of current-generation detectors. Having prior knowledge of the black hole's intrinsic parameters (mass and spin) and extrinsic parameters (luminosity distance and orientation) allows us to accurately predict the strain amplitude emitted by the source for a given $\alpha$ and thereby place confident constraints on the boson mass. 
An accurate estimate of the black hole age enables us to predict the optimal starting time of the search for a range of boson masses.
Prior knowledge of the sky position is also useful (in most cases; see Sec.~\ref{sec:sky_grid}), motivating us to target known, well-localized black holes constrained within $\sim 10^2$--$10^3$ deg$^2$. 

We can infer the parameters of a CBC remnant black hole from the inspiral-merger-ringdown signal observed by the detectors. The intrinsic and extrinsic source parameters listed in the previous paragraph are provided by CBC parameter estimation. We can pinpoint the time required for the cloud to grow and emit (if the corresponding boson particle exists) accurately since we know when the black hole was born. In addition, for sources seen by multiple detectors, we often have decent sky localization, particularly if there is an electromagnetic counterpart (i.e., at least one of the merging objects is a neutron star).


Reference~\cite{Isi2019} thoroughly discusses the benefits and drawbacks of targeting CBC remnants, as well as another potentially interesting target source: black holes in x-ray binaries. 
In this study, we show that, for vector boson signals, we are able to reach a much farther distance compared to scalar boson signals, and that current-generation detectors are capable of reaching sources at luminosity distances in line with some typical CBC remnants (Sec.~\ref{sec:sensitivity}). Thus, nearby CBC remnants are arguably the more desirable choice given the uncertainties associated with x-ray binary systems, and the fact that they will typically be much older.
Given that we do not have prior knowledge of the conjectured particle mass, it is in our best interest to target all black holes with reasonable potential to produce a detectable signal regardless of where they lie in the mass-spin plane. Targeting multiple black holes with different properties allows us to probe a larger boson mass range.
The fourth observing run of the LVK network is about to start with upgraded detectors, so we expect to have many remnant black holes suitable for vector boson studies.


\subsection{Sky localization uncertainty}
\label{sec:sky_grid}

In this section, we discuss in more detail how the sky localization of a CBC event would impact a vector boson search, and we analyze two different systems as examples.

For CBC remnant black holes detected by LIGO and Virgo, the sky positions are usually constrained to $\sim 10^1$--$10^3$~deg$^2$. 
When targeting a particular black hole, we need to run the search multiple times on a grid of sky positions to tile the patch in the sky where the source is believed to lie. We call these tiles ``sky templates." To minimize the computational cost, but also ensure we do not miss the signal, we must choose the number $N_{\rm sky}$ and spacing $\Omega_{\rm sky}$ of the sky templates carefully. This depends on a few factors: the position of the black hole, the size of the sky area constrained by the parameter estimation of the CBC signal, the signal strength, and the frequency resolution of the search.

A general guideline for selecting sky templates is to calculate the mismatch~\cite{Brady1998, Sun2018}. However, given the wide parameter space that needs to be covered in vector boson searches and the variety of search configurations required, more careful empirical verification is needed. 
Here we outline how to determine $N_{\rm sky}$ and $\Omega_{\rm sky}$ in a real search by injecting a synthetic signal into Gaussian noise ($S_h^{1/2} = 4 \times 10^{-24}$~Hz$^{-1/2}$) and searching over a grid of sky positions around the injection position. 
The signal strength is selected to be marginally above the detection threshold to ensure the search does not miss a weak signal due to a coarse sky grid. We investigate both a short-duration signal $\mathcal{O}$(hours) and a long-duration signal $\mathcal{O}$(months). We inject both signals at an arbitrarily chosen sky position RA = 20~hr and Dec = 10~deg, and we search over a $13 \times 13$ sky grid centered on the injection position. Table~\ref{tab:sky_grid} lists the detailed injection and search parameters for both signals.

\begin{table*}[tbh]
	\centering
	\setlength{\tabcolsep}{12pt}
	\renewcommand\arraystretch{1.2}
	\begin{tabular}{lccccccc}
		\hline
		Panel & $M_i~[M_{\odot}]$ & $\chi_i$ & $d_H$~[Mpc] & $\hpeak$ & Freq. band~[Hz] & $T_{\rm coh}$ & $T_{\rm obs}$ \\
		\hline
        Left & 60 & 0.7 & 600 & $4.41 \times 10^{-25}$ & 165--166 & 11.8~min & 18.1~hr \\
        Right & 100 & 0.5 & 800 & $5.19 \times 10^{-26}$ & 62--63 & 10.9~hr & 167.5~d \\
		\hline
	\end{tabular}
	\caption{Injection and search parameters used in Fig.~\ref{fig:sky_grid}.}
	\label{tab:sky_grid}
\end{table*}

\begin{figure*}[hbt!]
	\centering
	\includegraphics[scale=.6]{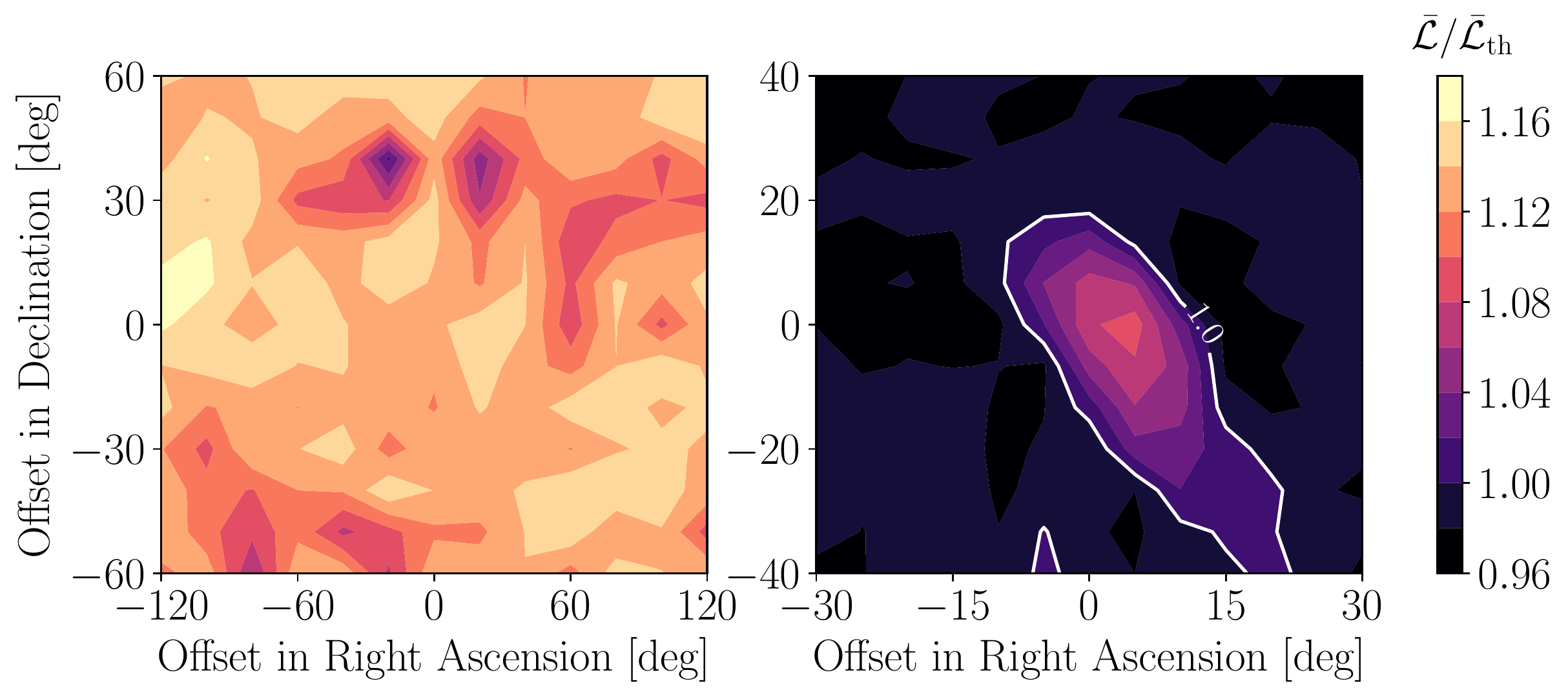}
	\caption{Colored contour of $\bar{\mathcal{L}}/\bar{\mathcal{L}}_{\rm th}$ as a function of the offset in RA and Dec for a short-duration signal (left) and a long-duration signal (right). See Table~\ref{tab:sky_grid} for the injection parameters. 
    Left: We have $\mathcal{L} > \bar{\mathcal{L}}_{\rm th}$ in the whole panel. Right: The bright EPSF enclosed within the white contour marks the region of the sky with $\bar{\mathcal{L}} > \bar{\mathcal{L}}_{\rm th}$ where the signal is successfully recovered.}
	\label{fig:sky_grid}
\end{figure*}

The results are shown in Fig.~\ref{fig:sky_grid}, with $\bar{\mathcal{L}}/\bar{\mathcal{L}}_{\rm th}$ evaluated at each sky position plotted as colored contours. The left and right panels show the short- and long-duration injections, respectively. 
The white contour in the right panel marks where $\bar{\mathcal{L}} = \bar{\mathcal{L}}_{\rm th}$, below which we do not recover the signal. We call the bright, above-threshold region the effective point spread function (EPSF) of the signal~\cite{Jones2022}. In the left panel, we have $\bar{\mathcal{L}} > \bar{\mathcal{L}}_{\rm th}$ over the whole grid.

In the right panel, the EPSF is slightly off-center, i.e., the maximum $\bar{\mathcal{L}}$ is not found at the injection position. 
The EPSF spans a large fraction of the sky, $\sim 20^2$~deg$^2$. 
This is the expected behavior for a search with $T_{\rm coh} \lesssim 1$~day, which has poor sky resolution~\cite{Jones2022}. 
If we set $\Omega_{\rm sky} \approx 20^2$~deg$^2$, we would only need $N_{\rm sky} \sim \mathcal{O} (1)$--$\mathcal{O} (10)$ to cover the relevant sky patch for a search with $T_{\rm coh} \lesssim 1$~day, assuming the source is reasonably well localized within $\sim 10^2$--$10^3$~deg$^2$,
which is easily attainable for an event seen by multiple detectors.
In the left panel, we do not see a clear EPSF because the signal occurs over a much shorter timescale with $T_{\rm coh}$ on the order of minutes. 
This is expected for very short-duration signals. We use the estimated sky position of the source to correct for the Doppler modulation due to the Earth's motion in the search. Because the modulation changes little over coherent integration times of $\sim 10$~min, the detection statistic is generally insensitive to offsets from the true sky position.
While such low sky resolution does come at the expense of degraded sensitivity, the major benefit here is that we do not need a sky grid to follow up a signal with $\tau_{\rm GW} \lesssim 1$~day.
As demonstrated in the above examples, the follow-up search for vector bosons targeting a CBC remnant black hole should be computationally practical, especially given the efficiency of the Viterbi algorithm.

\section{Conclusions}
\label{sec:conclusions}

In this paper, we explore how GW detectors can be used to uncover evidence of ultralight vector bosons through a process known as black hole superradiance. We implement a search technique for vector bosons around known black holes based on an HMM tracking scheme similar to the one used in directed searches for scalar bosons, but with certain modifications necessary to deal with the more rapidly evolving signals. 
We utilize a recently developed waveform model \texttt{SuperRad} to simulate GW signals from vector boson clouds, which allows us to optimize the search configuration and more accurately estimate its sensitivity.

In this study, we do not take into account any potential impact of the uncertainty in the signal waveform model on the search configuration.
The methods used here are much more flexible and less susceptible to model errors compared to, e.g., matched filter techniques.
Nevertheless, an overestimated first time derivative of the emitted GW frequency $\dot{f}_{\rm GW}$ may lead to a less optimal configuration (shorter $T_{\rm coh}$) for the search. An underestimated $\dot{f}_{\rm GW}$ may result in a loss of signal power as, at earlier times, the signal evolves more quickly than HMM can track.
Although we do not expect this to have any significant impact on the results here, future analyses may factor in the uncertainty estimate in the waveform model when available. Future improvements in the accuracy of the model may also be used to more finely tune the parameters of the search to their optimal values.

The computing cost for a given system depends on the parameter space to be covered and the signal duration, but is generally efficient. 
For instance, we can track a short-duration signal with $\tau_{\rm GW} \sim$ days in $\mathcal{O}$(10~min), whereas a typical long-duration signal with $\tau_{\rm GW} \sim$ months would take $\mathcal{O}$(1~hr) on a single core computer. 
A detailed scaling of computing cost as a function of $T_{\rm coh}$ and $T_{\rm obs}$ can be found in Ref.~\cite{Sun2018}.

We find that current-generation detectors can reach vector boson clouds at $\mathcal{O}$(1~Gpc) with our search methods for astrophysical black holes with $60~M_{\odot} \lesssim M_i \lesssim 600~M_{\odot}$ and $\chi_i \gtrsim 0.6$, corresponding to the boson mass $\sim 10^{-13}$~eV (see Fig.~\ref{fig:horizon_distances}). 
All CBC events detected by the first three LVK observing runs were within $\sim 5$~Gpc, with many detected at distances $\lesssim 1$~Gpc~\cite{CBCs_O1O2, CBCs_O3a, CBCs_O3b}. We expect more events like this with the upcoming fourth observing run. 
We also find that these searches are largely unimpacted by uncertainties in the sky position (see Sec.~\ref{sec:sky_grid}), making them even more practical.
Search plans are being made to follow up on promising CBC events.
Future-generation detectors, in addition to enabling searches for scalar bosons, will extend the reachable parameter space for vector bosons to nearly all CBC remnant black holes that are detected. 

\begin{acknowledgments}
We thank Max Isi for the helpful discussions and comments. 
DJ, LS, SS, and KW acknowledge the support of the Australian Research Council Centre of Excellence for Gravitational Wave Discovery (OzGrav), Project No. CE170100004. 
NS and WE acknowledge support from an NSERC Discovery grant. Research at Perimeter Institute is supported in part by the Government of Canada through the Department of Innovation, Science and Economic Development Canada and by the Province of Ontario through the Ministry of Colleges and Universities. This research was undertaken thanks in part to funding from the Canada First Research Excellence Fund through the Arthur B. McDonald Canadian Astroparticle Physics Research Institute.
The authors are grateful for computational resources provided by the LIGO Laboratory and supported by National Science Foundation Grants PHY--0757058 and PHY--0823459. 
This manuscript carries LIGO Document No. DCC--P2300081.
\end{acknowledgments}

\twocolumngrid

\end{document}